\documentclass[12pt,showkeys,amsmath,amssymb]{revtex4}
\usepackage{amsmath,amsfonts,amsthm,amscd,amssymb,latexsym}
\usepackage{bm}
\usepackage{dcolumn}
\usepackage{graphicx}
\usepackage{epstopdf}
\usepackage{color}
\usepackage{hyperref}
\usepackage{epsf}
\usepackage{epsfig}
\usepackage{graphicx, epic, eepic, color}
\usepackage{soul}
\usepackage{tikz}
\usetikzlibrary{calc}
\usetikzlibrary{decorations.markings}
\usetikzlibrary{shadings, calc, decorations.markings}

\def\beq{\begin{equation}}
\def\eeq{\end{equation}}

\begin{document}

\title{Characteristics of Effective Dark Matter in Nonlocal Gravity}

\author{Mahmood \surname{Roshan}$^{1,2}$}
\email{mroshan@um.ac.ir}
\author{Bahram \surname{Mashhoon}$^{2,3}$}
\email{mashhoonb@missouri.edu}

\affiliation{$^1$Department of Physics, Faculty of Science, Ferdowsi University of Mashhad, P.O. Box 1436, Mashhad, Iran \\
$^2$School of Astronomy, Institute for Research in Fundamental Sciences (IPM), P. O. Box 19395-5531, Tehran, Iran\\
$^3$Department of Physics and Astronomy, University of Missouri, Columbia, Missouri 65211, USA\\
}

\begin{abstract}
Nonlocal gravity (NLG) is a classical nonlocal generalization of Einstein's theory of gravitation that has been constructed in close analogy with the nonlocal electrodynamics of media. According to NLG,  what appears as dark matter in astrophysics and cosmology is in reality the nonlocal aspect of the universal gravitational interaction. We focus here on two main features of the effective dark matter in NLG, namely, (a) the density of effective dark matter in NLG is always finite and therefore cusp-free, and (b) there is less effective dark matter in dwarf galaxies than is generally assumed in the standard particle dark matter paradigm. The corresponding astrophysical implications of NLG in connection with three ultra-diffuse galaxies AGC 114905, 242019 and 219533 are discussed. 
\end{abstract}


\maketitle

\section{Introduction}

In 1933, Fritz Zwicky~\cite{FZ} applied the virial theorem of classical statistical mechanics to the Coma cluster of galaxies and concluded that the amount of mass within the cluster was not sufficient to hold the cluster together. He therefore suggested the possible existence of ``dark matter" within the cluster. The fundamental discovery of ``flat" rotation curves of spiral galaxies half a century ago provided observational proof  that something essential was missing in the standard description of dynamics of galaxies. To keep the longstanding  Kepler-Newton-Einstein tradition, one must postulate the existence of particle dark matter, which is needed to explain the dynamics of galaxies, clusters of galaxies and structure formation in cosmology. The properties of the hypothetical dark matter have thus far been deduced only through its gravity. The persistent negative result of experiments that have searched for the particles of dark matter has led to significant interest in theories that can explain away dark matter as an aspect of the universal gravitational interaction. That is, it is possible that there is no dark matter at all and the theory of gravitation needs to be modified on the scale of galaxies and beyond in order to take due account of what appears as dark matter in astrophysics and cosmology. The resulting gravitational \emph{effective dark matter} generated by a suitably extended theory of gravitation might then be able to account for observational data without any need for particle dark matter. 

The purpose of this paper is to discuss the nature of the effective dark matter in nonlocal gravity (NLG) theory, which is a classical nonlocal generalization of Einstein's theory of gravitation in which the nonlocal aspect of the universal gravitational interaction simulates dark matter. In classical field theory, electrodynamics of media is rendered nonlocal via the attributes of the background medium; that is, nonlocality enters the constitutive relations corresponding to Maxwell's original field equations~\cite{Jackson, L+L, HeOb}. It appears natural to consider a similar extension of Einstein's general relativity (GR) theory~\cite{Einstein}. The physical motivations for developing such a nonlocal gravity (NLG) theory as well as a comprehensive account of NLG is contained in~\cite{BMB}. To extend GR along the lines of the nonlocal electrodynamics of media, the initial step would involve expressing GR in a form that resembles Maxwell's electrodynamics. Indeed, there is a well-known teleparallel equivalent of general relativity (TEGR), which is the gauge theory of the group of spacetime translations~\cite{Cho}. Therefore, TEGR, though nonlinear, is formally analogous to electrodynamics and can be rendered nonlocal via history-dependent constitutive relations as in the nonlocal electrodynamics of media. In the resulting NLG theory~\cite{Hehl:2008eu, Hehl:2009es}, the gravitational field is locally defined but satisfies partial integro-differential field equations. The nonlocal theory employs an extended geometric framework that is consistent with universality of gravitational interaction and involves the standard spacetime curvature as in GR as well as spacetime torsion associated with a preferred frame field. 

Nonlocal gravity involves a certain average of the gravitational field over past events. The memory of the past appears in the gravitational field equation of NLG via a nonlocal retarded kernel. It appears that the field equation of NLG cannot be derived from a Lagrangian. That is, when one starts from a \emph{causal} Lagrangian density for linearized NLG involving a nonlocal kernel that is not symmetric in time, the variation of the corresponding action results in a gravitational field equation that contains a new kernel that is  \emph{symmetric} in time and thereby violates causality. Indeed, the new time-symmetric kernel is the sum of the original retarded kernel plus its corresponding advanced kernel, see~\cite{Hehl:2009es} for a detailed derivation of this result, which is consistent with the circumstance that there is no Lagrangian  for the  nonlocal electrodynamics of media.      

The only known exact solution of NLG is the trivial solution, namely, Minkowski spacetime in the absence of gravity. Thus far, the nonlinearity of NLG has prevented finding exact solutions for strong-field regimes such as those involving black holes or cosmological models~\cite{Bini:2016phe}. However, linearized NLG and its Newtonian limit have been extensively studied~\cite{BMB}. 

When we write the field equations of NLG in the same form as GR field equations,  we find that besides the standard symmetric energy-momentum tensor of matter, certain purely nonlocal gravity terms appear which we can interpret in terms of nonlocally induced effective dark matter. What is now considered dark matter in astrophysics and cosmology may indeed be the manifestation of the nonlocal component of the universal gravitational interaction.

In view of the possible astrophysical applications of NLG, in this paper we focus on the properties of the effective dark matter in the Newtonian regime of NLG. In this limit,  the gravitational force on a test particle of inertial mass $m$ is given by
\begin{equation}\label{I0}
\mathbf{F}(\mathbf{x}) = - m \nabla\Phi(\mathbf{x})\,,
\end{equation}
where the gravitational potential $\Phi(\mathbf{x})$ satisfies 
\begin{equation}\label{I1}
\nabla^2\Phi (\mathbf{x}) + \int \eta(\mathbf{x}-\mathbf{y}) \nabla^2\Phi (\mathbf{y})\,d^3y = 4\pi G\,\rho (\mathbf{x})\,.
\end{equation}
Here,  $\eta$ is the universal kernel of NLG in the Newtonian regime. Let us assume that we work in the space of functions that are absolutely integrable ($L^1$) as well as square integrable ($L^2$); then,  it is possible to write Eq.~\eqref{I1} in its reciprocal form 
\begin{equation}\label{I2}
4\pi G\,\rho (\mathbf{x}) +  \int q(\mathbf{x}-\mathbf{y}) [4\pi G\, \rho(\mathbf{y})]\,d^3y = \nabla^2\Phi (\mathbf{x})\,,
\end{equation}
where $q$ is the reciprocal kernel~\cite{Chicone:2011me, BMB}. It is possible to write the nonlocal Poisson Eq.~\eqref{I2} in the form
\begin{equation}\label{I3}
\nabla^2\Phi = 4\pi G\,(\rho+\rho_D)\,,
\end{equation}
where the origin of $\rho_D$ is the nonlocal aspect of gravity that appears as an extra source of matter. We interpret $\rho_D$ as the density of effective dark matter given by the convolution of the reciprocal kernel $q$ with the density of matter $\rho$, namely, 
\begin{equation}\label{I4}
\rho_D(\mathbf{x})=\int q(\mathbf{x}-\mathbf{y}) \rho(\mathbf{y})\,d^3y\,.
\end{equation}
Thus, nonlocality appears to simulate dark matter. In principle, $\Phi$, $\rho$ and $\rho_D$ could depend upon time $t$; however, this possibility has been suppressed here for the sake of simplicity. The determination of
 $\rho_D(\mathbf{x})$ as a convolution defined by Eq.~\eqref{I4} is a significant result of the Newtonian regime of NLG. Let us observe that $\rho_D = 0$ when $\rho = 0$; that is, there is no effective dark matter in the complete absence of matter. \emph{In NLG, effective dark matter shadows matter.} 

In the nonlocal electrodynamics of media, the kernel is determined on the basis of the atomic physics of the underlying medium; however, the situation is clearly different in nonlocal gravity. In the case under consideration here, we must determine the reciprocal kernel $q$ on the basis of observational data. Moreover, $q$ must be absolutely integrable as well as square integrable. As described in detail in~\cite{BMB}, NLG in the Newtonian regime recovers the phenomenological Tohline-Kuhn approach to modified gravity~\cite{Toh,Kuhn,Bek}, where the spherically symmetric Kuhn kernel 
$(4\pi \lambda_0\,r^2)^{-1}$ had been determined in accordance with the flat rotation curves of spiral galaxies. Here, $\lambda_0 \sim 1$~kpc is the basic galactic length scale in this approach. Introducing two other length scales $a_0$ and $1/\mu_0$ to moderate the short and long distance behaviors of $q$, respectively, two simple spherically symmetric $L^1$ and $L^2$ functions have been studied in detail as generalizations of Kuhn's kernel, namely, 
\begin{equation}\label{I5}
 q_1 (r) = \frac{1}{4\pi \lambda_0} \,\frac{1+\mu_0\, (a_0+r)}{r\,(a_0 + r)}\,e^{-\mu_0\,r}\,, \qquad q_2 (r) = \frac{r}{a_0 + r}\,q_1(r)\,,
\end{equation}  
where $r = |\mathbf{x} - \mathbf{y}|$ and Kuhn's kernel is recovered for $a_0 = \mu_0 = 0$. We note that nonlocality disappears when $\lambda_0$ tends to infinity; that is, the Tohline-Kuhn  parameter  
$\lambda_0$ is the essential nonlocality parameter of the Newtonian regime of NLG. Moreover, if $\rho$ is spherically symmetric, then so is $\rho_D$ due to the assumed spherical symmetry of $q$. We emphasize that the form of the reciprocal kernel is in no way unique and can be significantly modified if required by observational data. 

Solar system data provide a lower bound for $a_0$, namely, $a_0 \gtrsim 10^{15}$~cm~\cite{Chicone:2015coa}. It is interesting to note that 
\begin{equation}\label{I6} 
\frac{\partial q_1}{\partial a_0} <0 \,, \qquad \frac{\partial q_2}{\partial a_0} <0\,,
\end{equation}
which implies that $\partial \rho_D / \partial a_0 <0$, i.e. with increasing $a_0$, the effective dark matter density decreases. It seems that finding the appropriate value of $a_0$ using galactic data is not so simple; that is,  rotation data do not extend inward sufficiently close to the centers of spiral galaxies.   On the other hand, if we ignore $a_0$, i.e. with $a_0 = 0$, $q_1 = q_2 = q_0$,
\begin{equation}\label{I7}
 q_0 (r) = \frac{1}{4\pi \lambda_0} \,\frac{1+\mu_0\,r}{r^2}\,e^{-\mu_0\,r}\,,
\end{equation}
where for any finite radial coordinate $r$,  $q_0 > q_1 > q_2$, since $a_0 > 0$. It proves useful to define dimensionless parameters $\alpha_0$ and $\zeta_0$, 
\begin{equation}\label{I8}
\alpha_0 := 2/(\lambda_0\,\mu_0)\,, \qquad \zeta_0 := a_0\mu_0\,.
\end{equation}  
The rotation curves of nearby spiral galaxies can be used to find $\alpha_0$ and $\mu_0$. Indeed, observational data regarding nearby spiral galaxies and clusters of galaxies are consistent 
with~\cite{Rahvar:2014yta}
\begin{equation}\label{I9}
\alpha_0 = 10.94 \pm 2.56\,,\quad \mu_0 = 0.059 \pm 0.028~{\rm kpc}^{-1}\,, \quad  \lambda_0 = \frac{2}{\alpha_0\,\mu_0} \approx 3\pm 2~{\rm kpc}\,.
\end{equation}

The density of effective dark matter $\rho_D$ is the convolution of the density of matter $\rho$ with the empirically determined reciprocal kernel $q$ of NLG. We naturally assume throughout that 
$\rho(\mathbf{x})$ is a positive function that is both integrable as well as square integrable. The notion of convolution (folding) has interesting properties in the Fourier domain. Let $\hat{s} (\boldsymbol{\xi})$ be the Fourier integral transform of a function $s(\mathbf{x})$ that is both $L^1$ and $L^2$; then, 
\begin{equation}\label{I10}
\hat{s} (\boldsymbol{\xi}) =  \int s(\mathbf{x})\, e^{-i\,\boldsymbol{\xi} \cdot \mathbf{x}}\, d^3x, \qquad 
s(\mathbf{x})=\frac{1}{(2\pi)^3}\,\int \hat{s} (\boldsymbol{\xi})\,e^{i\,\boldsymbol{\xi} \cdot \mathbf{x}}\, d^3\xi.
\end{equation}
In our case, we can write
\begin{equation}\label{I11}
\hat{\rho}(\mathbf{k}) = \int \rho(\mathbf{x})\,e^{-i\,\mathbf{k}\cdot \mathbf{x}}\,d^3x\,\,, \qquad \rho(\mathbf{x}) = \frac{1}{(2\pi)^3}\,\int \hat{\rho}(\mathbf{k})\,e^{i\,\mathbf{k}\cdot \mathbf{x}}\,d^3k\,
\end{equation}
and so on for $\rho_D$ and $q$. Then, the convolution theorem states
\begin{equation}\label{I12}
\hat{\rho}_D(\mathbf{k}) = \hat{q}(\mathbf{k})\, \hat{\rho}(\mathbf{k})\,.
\end{equation}

Let us note that $\hat{\rho}(0)$ in Eq.~\eqref{I11} is equal to the total mass of the source $M$; that is, 
\begin{equation}\label{Ia}
\hat{\rho}(0) = \int \rho(\mathbf{x})\,d^3x\, = M\,.
\end{equation}
This is a general result and applies to other Fourier integral transforms as well. Therefore, evaluating Eq.~\eqref{I12} at $\mathbf{k} = 0$, we find the important relation
\begin{equation}\label{I13}
M_D = \int \rho_D(\mathbf{x}) d^3x =   M \,\int q(\mathbf{x}) \,d^3x\,.
\end{equation}
Here, 
\begin{equation}\label{I14}
\int q_i(\mathbf{x})\,d^3x = \alpha_0\, w_i\,, \quad w_1= 1- \frac{1}{2}\zeta_0\,e^{\zeta_0} E_1(\zeta_0)\,, \quad w_2 = 1- \zeta_0\,e^{\zeta_0} E_1(\zeta_0)\,,
\end{equation}
where $a_0\mu_0 = \zeta_0$ and $E_1(x)$ is the \emph{exponential integral function}~\cite{A+S}. It follows that 
\begin{equation}\label{I15}
M_D =  \alpha_0\, w\,M\,,
\end{equation}
where $w$ is equal to $w_1$ or $w_2$ depending upon whether the reciprocal kernel is $q_1$ ior $q_2$, respectively. For most physical applications, we expect that $\zeta_0 \ll 1$  and one can in effect put $a_0 = 0$. In this case, considerable simplification occurs as the reciprocal kernel reduces to $q_0$ given by Eq.~\eqref{I7}. In particular, $M_D = \alpha_0 M$ in Eq.~\eqref{I15}. The integral of $q_0$ over all space is $\alpha_0$  and its Fourier integral transform is given by
\begin{equation}\label{I16}
\hat{q}_0(\mathbf{k}) = \alpha_0\,Q (|\mathbf{k}|/\mu_0)\,, \qquad   Q(u) =  \frac{1}{2} \left(\frac{1}{1 + u^2} +\frac{1}{u}\,\arctan{u}\right)\,,
\end{equation}
where $Q(u)$ is a positive function that starts from unity at $u = 0$, decreases monotonically  and vanishes as $u^{-2}$ for $u \to \infty$. Let us recall that $q$ has been constructed to be an integrable as well as a square-integrable function; on the other hand, if $a_0 = 0$, then $q_1 =q_2 = q_0$ given by Eq.~\eqref{I7} is integrable but not square integrable. 

In Section II, we show that in NLG the effective dark matter density profile is cusp-free; that is, $\rho_D(\mathbf{x})$ is always finite. In Section III, we study $\rho_D(r)$ throughout space when the density of matter is given by the Plummer model~\cite{Plummer}. We discuss the dependence of $M_D$ upon the size of the galactic system in Section IV. Section V is devoted to ultra-diffuse galaxies (UDGs), namely,  AGC 114905, 242019 and 219533. We summarize our results in the final Section VI. 

\section{$\rho_D$ is Cusp-Free}

The density of the effective dark matter in NLG, namely, $\rho_D$ is the convolution of the matter density $\rho$ and the reciprocal kernel $q$. The functions $\rho$ and $q$ are both positive by definition. They are also integrable functions. It is a mathematical theorem that under these conditions their convolution $\rho_D$ is finite and positive; in particular,  $\rho_D$ is cusp-free. Moreover, since the reciprocal kernel $q$ is integrable as well as square integrable, $\rho_D$ is integrable as well as square integrable by Young's inequality for convolutions~\cite{Bog}. In some computations of  $\rho_D$, it is simpler to use $q_0$ instead of $q$. Indeed, the special kernel $q_0$ is integrable; therefore, we can safely use it instead of $q$ in Eq.~\eqref{I4} to calculate $\rho_D$, once ignoring $a_0$ can be physically justified. 

There is a cusp-core problem in connection with the nature of dark matter halos within the framework of cold dark matter (CDM) cosmology~\cite{Binney}. For dwarf spheroidal galaxies, $N$-body simulations generally predict  ``cuspy" dark matter distributions; that is,  the density of dark matter increases steeply at small radii. On the other hand, observational data regarding the rotation curves of most dwarf galaxies suggest a constant central dark matter density profile; moreover, such cored profiles are consistent with corresponding dynamical models. For a recent discussion of the problem, see~\cite{Boldrini:2019yvk} and the references cited therein. An example of a cuspy dark matter profile is the Navarro-Frenk-White spherical density distribution~\cite{NFW} 
\begin{equation}\label{M1} 
\rho_{\rm NFW} (r) = \rho_s\,\frac{R_s}{r\,(1+ r/R_s)^2}\,,
\end{equation}
where $r$ is the radial coordinate, $\rho_s$ is a constant density and $R_s$ is some scale radius. Thus, $\rho_{\rm NFW} (r) \propto r^{-1}$ diverges at the center. 

The density of effective dark matter in NLG is cusp-free and is thus expected to have a cored profile. Such profiles have been suggested for dwarf spheroidal galaxies (dSphs). For instance, it is claimed in~\cite{universal} that the dark matter density of dSphs can be universally modeled by the following cored profile:
\begin{equation}\label{M2} 
\rho_d(r)=\rho_h\left(1+\frac{r}{r_h}\right)^{-3}\,,
\end{equation}
where $\rho_h$ and $r_h$ are constants. In the process of comparison with the observational data, the stellar mass-to-light ratio $\Upsilon_{*}$ is not required in this case, but it would be necessary for NLG as $\rho_D$ depends on the density of matter in the corresponding dSph. Observational data for dSphs do not appear to be robust enough at present to allow a fair comparison of $\rho_D$ of NLG in this case with the standard dark matter cored models. In this connection, we note that at present cuspy NFW halo models can sometimes also work for dSphs (as the second universal model).  It appears that the baryonic observational features can sometimes be modeled with two halo models that behave so differently at $r=0$. This only means that observations (such as dispersion velocities) are not yet sensitive enough in all cases to recognize the slope of the dark matter density at $r=0$.

It is interesting to work out the effective dark matter distribution in a specific case. The baryonic matter density is typically modeled by the Plummer profile in essentially all of the dSphs~\cite{universal}. This circumstance provides the motivation to study $\rho_D$ in dSphs using the Plummer model.

\section{Cored Profile of $\rho_D(r)$ in the Plummer Model}

We are interested in $\rho_D$ given by Eq.~\eqref{I4} when $\rho$ is given throughout all space by the Plummer model~\cite{Plummer}, 
\begin{equation}\label{P1} 
\rho_{\rm P}(r) =\ \frac{\rho_0}{(1 + r^2/R_0^2)^{5/2}}\,,
\end{equation}
so that there is no outer boundary for the matter distribution. Here $\rho_0$ and $R_0$ are constants. Let $\mathbf{x}-\mathbf{y} := \mathbf{z}$; then, Eq.~\eqref{I4} can be written as 
\begin{equation}\label{P2} 
\rho_D(\mathbf{x})=\int q(\mathbf{z}) \rho_{\rm P}(\mathbf{x} -\mathbf{z})\,d^3z\,.
\end{equation}
Hence, 
\begin{equation}\label{P2a} 
\rho_D(\mathbf{x}) = 2 \pi \rho_0 \int_{0}^{\infty} q(z) z^2 dz \int_{-1}^{+1}\frac{d\Theta}{[1+(r^2+z^2 -2rz \Theta)/R_0^2]^{5/2}}\,,
\end{equation}
where $r = |\mathbf{x}|$, $z = |\mathbf{z}|$ and $\Theta := \cos\theta$. The $\Theta$ integration can be simply carried out and the result is
\begin{equation}\label{P3} 
\rho_D(r)= \frac{2 \pi\, \rho_0 R_0^5}{3 r} \int_{0}^{\infty} q(z) z dz \left\{[R_0^2 +(z-r)^2]^{-3/2} - [R_0^2 +(z+r)^2]^{-3/2}\right\}\,,
\end{equation}
which is spherically symmetric. It is indeed generally the case that if the matter distribution is spherically symmetric, the same will be true for the effective dark matter distribution due to the assumed spherical symmetry of the reciprocal kernel $q$.  

Let us note that as $R_0 \to \infty$, $\rho_{\rm P}(r) \to \rho_0$ and $\rho_D(r) / \rho_0$ approaches the integral of $q$ over all space, as expected. Moreover,  $\rho_D$ as a function of $r: -\infty \to \infty$ is an even function, i.e. $\rho_D(-r) = \rho_D(r)$, and therefore has an extremum at $r = 0$ with zero slope. It has indeed a maximum at $r=0$; that is, the Taylor expansion of $\rho_D(r)$ about $r = 0$ is given by
\begin{equation}\label{P4} 
\rho_D(r)= \rho_D(0) + \frac{1}{2} \rho_D'' (0) \,r^2 + O(r^4)\,,
\end{equation}
where 
\begin{equation}\label{P5} 
\rho_D(0)= 4 \pi \rho_0 \int_{0}^{\infty} \frac{q(z) z^2 dz}{(1+z^2/R_0^2)^{5/2}}\, 
\end{equation}
and
\begin{equation}\label{P6} 
\rho_D''(0)= -\frac{20 \pi \rho_0}{R_0^2} \int_{0}^{\infty}\left(1-\frac{4}{3}\frac{z^2}{R_0^2}\right) \frac{q(z) z^2 dz}{(1+z^2/R_0^2)^{9/2}}\,.
\end{equation}
It can be shown that this quantity, i.e. the second derivative of density of effective dark matter at the origin, is always negative regardless of the magnitude of $ \mu_0 R_0> 0$.

For $r \to \infty$, we find 
\begin{equation}\label{P7} 
\rho_D(r \to \infty)\sim 4 \pi \rho_0 \frac{R_0^5}{r^5} \int_{0}^{\infty} q(z) z^2 dz = \rho_{\rm P}(r \to \infty) \int q(z) d^3z\,,
\end{equation}
as expected. We recall that the integral of $q$ over all space is given by $\alpha_0\,w$, where $w$ is a positive constant such that $w = 1$ when $q (a_0 = 0)= q_0$. For $a_0 \ne 0$, $q$ can be either $q= q_1$ or $q=q_2$. The corresponding values of $w$ are $w_1$ or $w_2$ given by Eq.~\eqref{I9}. Indeed, $1-w_1$ and $1-w_2$ are both positive and depend only upon $\zeta_0 = a_0 \mu_0$. In Figure 1, we have used reciprocal kernel $q_1$ given in Eq.~\eqref{I5} to calculate $\rho_D$ given in the left panel, while the right panel contains the plots of the rotation curves $V_{\rm P}(r)$, $V_D(r)$ and 
$V = [V_{\rm P}^2(r) + V_D^2(r)]^{1/2}$ in this case; moreover,  $\mu_0 \,R_0= 0.04$ and $a_0= \,0.049 R_0$ in both panels. Here,
\begin{equation}\label{Pa} 
V_{\rm P}^2 = \frac{4 \pi G}{r}\int_0^r \rho_{\rm P} (s) s^2 \,ds\,, \qquad V_D^2 = \frac{4 \pi G}{r}\int_0^r \rho_D (s) s^2 \,ds\,, 
\end{equation}
as a direct consequence of Newton's shell theorem. 

Although the dispersion velocity ($\sigma$) is measured for dSphs, the rotation curve is still important. We note that it has been shown in a recent paper \cite{McGaugh:2021tyj}  that there is a close relationship between dispersion velocity and the corresponding rotation curve (obtained from mass distribution) in dSphs; more specifically, $V\approx 2 \sigma$. On the other hand, the effective dark matter density plot shows the absence of the cusp. This feature cannot be easily inferred from the rotation curve.

\begin{figure}
\centering
\includegraphics[width=7.0cm]{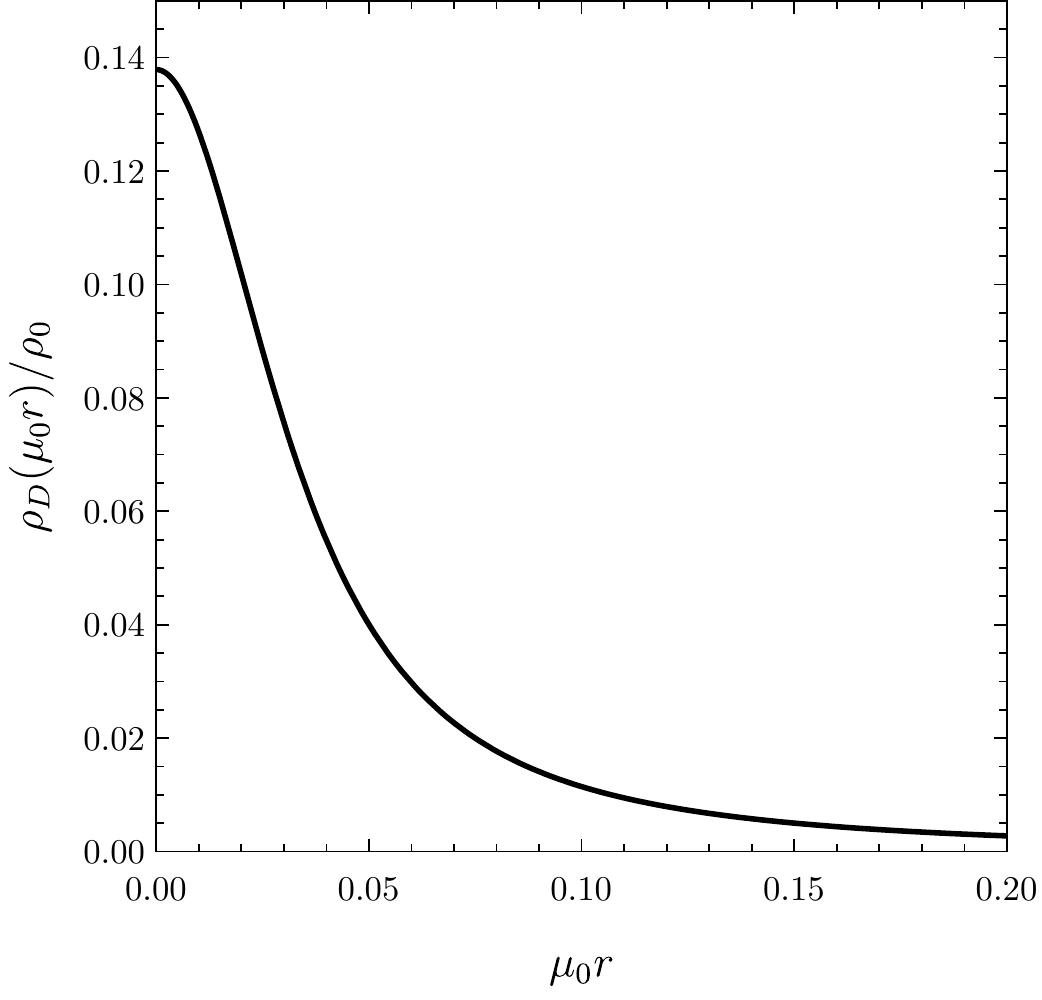}\hspace{0.2cm}
\includegraphics[width=7.1cm]{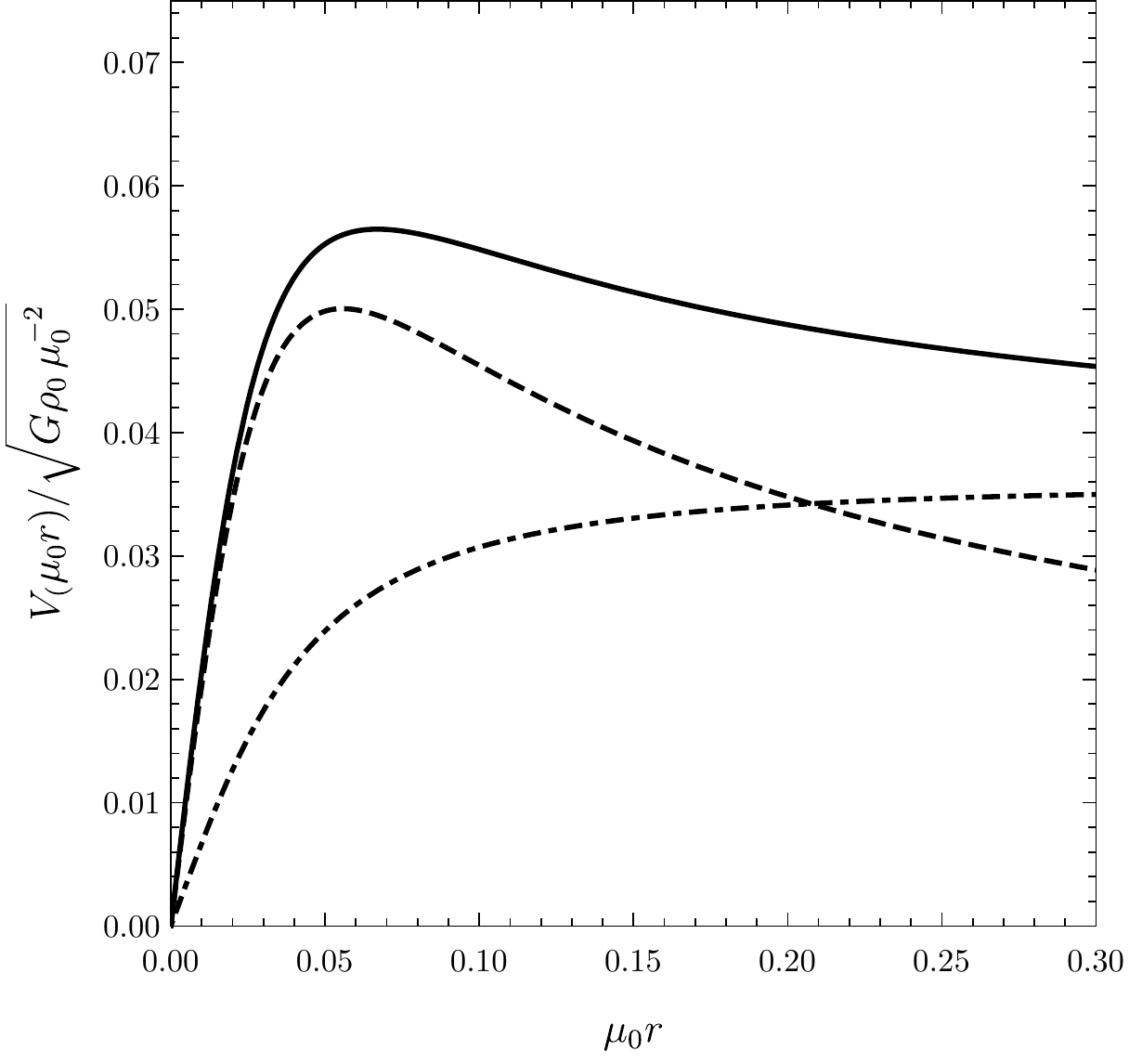}
\caption{\textit{Left panel}: Plot of $\rho_D/\rho_0$ for the Plummer model. Here, $\rho_D$ is the effective dark matter density and $\rho_0$ is the central density of matter for the Plummer sphere. \textit{Right panel}: The dot-dashed and dashed curves indicate the contributions of the effective dark matter density 
$\rho_D$ and the Plummer density $\rho_{\rm P}$, respectively, to the total rotation curve represented by the solid line. }\label{plumden}
\end{figure}

To emphasize this point, let us explore the implications of the presence of a cusp for the general rotation curve $V(r)$ as $r\to 0$. Here, we assume spherical symmetry as in Eq.~\eqref{Pa}.  If there is no cusp and the density is finite at $r = 0$, then $V(r)/r$ approaches a constant as $r \to 0$. This is the case in NLG. On the other hand, if in the standard dark matter paradigm the density of dark matter has a cusp and for $r\to 0$ behaves as $r^{-\varpi}$ with $2 > \varpi > 0$, then $V(r) \propto r^{1-\varpi/2}$ and $dV/dr$ diverges as $r\to 0$. For $\varpi = 2$, $V(r = 0)$ is a nonzero constant and for $\varpi > 2$, $V(r)$ diverges as $r\to 0$. As the collection of observational data for the rotation curve normally occurs at a considerable distance from the center, it is not possible to deduce the existence of a cusp in the central density of dark matter from the observed rotation curve.

Finally, let us mention that since the characteristic size of dSphs is much smaller than $\mu_0^{-1}\approx 17\,$kpc, one can use an expansion in powers of  $\mu_0 r \ll1$ to calculate analytically the effective dark matter density in this case.

\section{Effective Dark Matter Within a Galaxy}

Consider an astrophysical system that consists essentially of a large number of distinct classical point particles. The density of matter $\rho$ and the total mass of the system $M$ can be written as 
\begin{equation} \label{G1}
\rho(\mathbf{x}) = \sum_p m_p\, \delta(\mathbf{x} - \mathbf{x}_p)\,, \qquad M = \sum_p m_p\,.
\end{equation}
According to the nonlocal Poisson Eq.~\eqref{I3}, the gravitational potential of the system in NLG can be obtained from 
\begin{equation} \label{G2}
\nabla^2\Phi = 4\pi G\,\sum_p m_p\,[\delta(\mathbf{x} - \mathbf{x}_p) + q(\mathbf{x} - \mathbf{x}_p) ]\,.
\end{equation}
To the traditional Newtonian gravitational potential of the system, one must add the contribution of the effective dark matter which has the form of a spherical distribution centered on each point particle of mass $m_p$ with density $m_p\,q(r)$, where $r = |\mathbf{x} - \mathbf{x}_p|$ is the radial distance away from $m_p$. The nature of this distribution of effective dark matter is the same for each point particle and is given by the reciprocal kernel $q(r)$, which is a positive function that monotonically decreases with increasing radius $r$ as it decays exponentially with decay length $\mu_0^{-1} \approx 17$ kpc. This description is valid for $q_1(r)$ and $q_2(r)$ defined in Eq.~\eqref{I5} as well as for $q_0(r)$ defined in Eq.~\eqref{I7}. \emph{The classical point particle of Newtonian mechanics is replaced in NLG by the point particle together with a spherical cocoon of effective dark matter.}

Let us briefly digress here and illustrate the importance of the cocoon in NLG by calculating the gravitational force on a point particle of mass $m'$ located at $\mathbf{x}'$ due to the point mass $m_p$ at $\mathbf{x}_p$ and its associated cocoon. It follows from Eqs.~\eqref{I0} and~\eqref{G2} that the result is the sum of the standard Newtonian force due to $m_p$ and its cocoon, which acts like another point particle at $\mathbf{x}_p$. The force due to the cocoon, according to Newton's shell theorem, is Newtonian and involves in essence the new particle at $\mathbf{x}_p$ with a mass that is the total effective dark matter of density $m_p\, q$ within a sphere of radius $|\mathbf{x}' - \mathbf{x}_p|$; that is, 
\begin{equation}\label{G2a}
\mathbf{F}_{\rm NLG}(\mathbf{x}') =  Gm' m_p\,\frac{\mathbf{x}_p - \mathbf{x}'}{|\mathbf{x}_p - \mathbf{x}'|^3} \left[1+ \int_0^{|\mathbf{x}_p - \mathbf{x}'|} 4\,\pi\,s^2 q(s) ds\right]\,.
\end{equation}

For each point particle $m_p$, the mass of the cocoon, i.e. the net amount of effective dark matter throughout the universe, is given by 
\begin{equation} \label{G3}
 m_p\, \int_0^\infty 4 \pi r^2 q(r) dr = m_p \alpha_0 w\,, 
 \end{equation}
 in agreement with Eq.~\eqref{I15}. Here, $\alpha_0 \approx 11$ and $w$ is somewhat less than unity due to the existence of the short-distance parameter $a_0$ in the reciprocal kernel. For $a_0 = 0$, $q=q_0$ and $w=1$. We conclude that the mass of the cocoon associated with a particle in NLG is about an order of magnitude larger than its inertial mass. On the other hand, an astrophysical system is normally assigned a bounded region of space. Let $\mathcal{D}$ be the diameter of the smallest sphere that completely surrounds the astronomical system under consideration here. The boundary of the system, confined within the sphere of diameter $\mathcal{D}$, would naturally cut out and exclude the outer part of the cocoon associated with each point particle within the system. We are thus left with an amount of effective dark matter inside the system that is less than $M \alpha_0 w$, i.e. the sum of Eq.~\eqref{G3} over the system. However, if $\mathcal{D} \gg 17$ kpc, then most of the effective dark matter is captured within the system since the spherical distribution involves an exponential decay length of $\approx 17$ kpc. Therefore, we expect that $M_D \approx M \alpha_0 w$ for a giant galaxy or a cluster of galaxies. This order-of-magnitude result of NLG appears to be in general agreement with observational data regarding nearby giant galaxies and clusters of galaxies~\cite{BMB, Rahvar:2014yta}. 

\begin{figure}
\centering
\includegraphics[width=8.0cm]{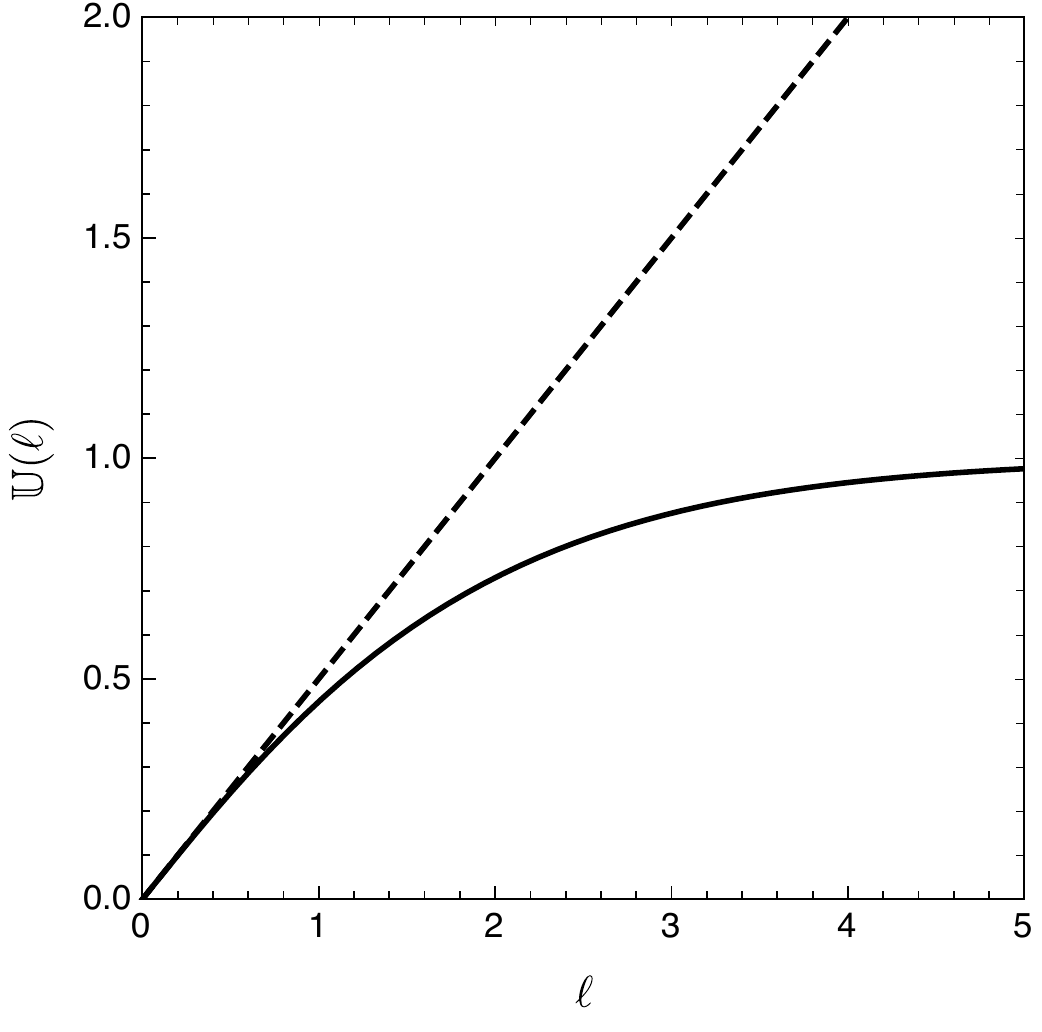}
\caption{The solid curve indicates the plot of $\mathbb{U}$ versus $\ell$. The dashed line is $\ell/2$.}\label{fig2}
\end{figure}

To estimate the net amount of effective dark matter within a system with radius comparable or less than $\mu_0^{-1} \approx 17$ kpc, imagine a point mass $m_p$ within the system and the surface that is the boundary of the cocoon and the system; then, the distance from $m_p$ to this surface, $r = |\mathbf{x} - \mathbf{x}_p|$,  can be at most $\mathcal{D}$. This means that the amount of effective dark matter associated with $m_p$ that remains within the system is less than
\begin{equation} \label{G4}
 m_p\, \int_0^{\mathcal{D}} 4 \pi r^2 q(r) dr\,. 
\end{equation}
Summing this result over the system, we naturally find
\begin{equation} \label{G5}
M_D < M\, \int_0^{\mathcal{D}} 4 \pi r^2 q(r) dr\,. 
\end{equation}
We emphasize that this simple result is based on a crude estimate and it would be desirable to have a better approximation for $M_D$ that we can then compare with observational data. In any case, we recall that $q < q_0$; therefore,  
\begin{equation} \label{G6}
M_D < M \int_0^{\mathcal{D}} 4 \pi r^2 q_0(r) dr = \frac{1}{2} \alpha_0 M \int_0^{\mu_0 \mathcal{D}} (1+x)e^{-x} dx\,, 
\end{equation}
where we have employed Eqs.~\eqref{I7} and~\eqref{I8}. Hence, 
\begin{equation} \label{G6}
M_D <   \alpha_0  M \, \mathbb{U}(\mu_0 \mathcal{D})\,, 
\end{equation}
where
\begin{equation} \label{G7}
\mathbb{U}(\ell) = 1 - (1 + \tfrac{1}{2}\ell) e^{-\ell}\,. 
\end{equation}
For $\ell: 0 \to \infty$, $\mathbb{U}(\ell)$ is a smooth type of a unit step function that starts from $0$ at  $\ell = 0$, monotonically increases with slope $1/2$ and asymptotically approaches unity as 
$\ell \to \infty$. As illustrated in Fig. \ref{fig2}, $\mathbb{U}(\ell)\le \ell/2$; that is,  for $\ell > 0$, $\mathbb{U}(\ell)$ always stays below the line $\ell/2$, which approximates  $\mathbb{U}(\ell)$ reasonably well
 for $\ell \lesssim1$. Putting these results together, we find
\begin{equation} \label{G8}
M_D <   M \, \frac{\mathcal{D}}{\lambda_0}\,, 
\end{equation}
where $\lambda_0 \approx 3$ kpc is the basic length scale of NLG  and we have used $\alpha_0\mu_0 = 2/\lambda_0$. 

Galactic systems for which $\mathcal{D}$ is comparable with $\mu_0^{-1} \approx 17$ kpc have been briefly considered in Ref.~\cite{BMB} and the amount of the effective dark matter is in reasonable agreement with observation. Furthermore, in the extreme case of a globular star cluster with its $\mathcal{D}$ about 40 pc, NLG predicts that the amount of effective dark matter is less than a few percent of the mass of the globular cluster. This circumstance appears to be consistent with the usual assumption that there is essentially no dark matter in globular clusters. On the other hand, for dwarf galaxies Eq.~\eqref{G8} predicts that the effective amount of dark matter is less than a quantity that is at most a few times the mass of the system. However, dwarf galaxies are generally assumed to be dark matter dominated systems. Therefore, NLG predicts less effective dark matter for dwarf galaxies as compared with the standard picture. This prediction seems to be consistent with observational data regarding dwarf galaxies that lack dark matter~\cite{vanDokkum:2018vup, Guo:2019wgb, Pina:2019rer, Hammer:2020qcd, Shen:2021zka}. In the next section, we apply NLG to three UDGs, namely, AGC 114905, 242019 and 219533  and compute their effective dark matter content. 

Finally, it is useful to mention some NLG predictions for the density of the effective dark matter $\rho_D$ within spiral galaxies and contrast them with expectations based on the standard cold dark matter paradigm. Indeed, some specific aspects of $\rho_D$  may make it distinguishable from the density of the standard cold dark matter. The effective dark matter in NLG tracks the distribution of baryonic matter. In the case of spiral galaxies, the spiral arms show up in mapping the distribution of the corresponding effective dark matter. This is generally not the case in the cold dark matter framework. Moreover, the effective dark matter distribution has essentially the same symmetry properties as the disk of the galaxy. This means that unlike in the standard picture, where cold dark matter halos are modeled by NFW spherical halos, the distribution of the effective dark matter is not spherical. This would directly affect the secular evolution of the galactic disk in vertical and radial directions~\citep{Roshan:2021mfc,Roshan:2021ljs}.

\section{Ultra-diffuse galaxies in NLG}

An UDG is a low-surface-brightness galaxy that has a larger effective radius than a dwarf galaxy with the same mass. Moreover, in gas-rich UDGs the baryonic to total mass fraction is much higher than that of normal galaxies with similar rotation curves. This directly means that UDGs contain less dark matter and do not respect the baryonic Tully-Fisher relation \cite{McGaugh:2000sr}. In recent  years,  some nearby  dark-matter-free UDG candidates have attracted much attention; for example, 
see~\cite{vanDokkum:2018vup, Guo:2019wgb, Pina:2019rer, Hammer:2020qcd, Shen:2021zka}. 

In the standard cold dark matter paradigm, one imagines that luminous galaxies are embedded within dark matter halos. In the process of galaxy formation within the potential well provided by dark matter, the total mass of the stars in the galaxy ($M_{\rm stars}$) is expected to become correlated with the corresponding halo mass ($M_{\rm halo}$). In this framework, the ratio of the halo mass to the mass of the stars, 
$M_{\rm halo}/M_{\rm stars}$, when plotted versus $M_{\rm stars}$, has a minimum value of around $30$ for the Milky Way type of galaxies and increases toward lower and higher galactic masses~\cite{vanDokkum:2018vup, Guo:2019wgb, Pina:2019rer, Hammer:2020qcd, Shen:2021zka}. 
For  the UDGs under consideration here the ratio $M_{\rm halo}/M_{\rm stars}$ is about unity; therefore, they  pose a significant challenge to the standard dark matter paradigm.  Nevertheless, it is necessary to
 mention that a scenario has recently been suggested for  the formation of UDGs. They might have been created during  collisions between dwarf galaxies where the baryonic matter can, in principle, be separated from the dark counterpart~\cite{vanDokkum:2022zdd}.

Nonlocal gravity theory naturally predicts less dark matter for dwarf galaxies. What are the implications of NLG for UDGs?  Although the rotation curves and dynamics of nearby normal spiral galaxies have already been investigated within the context of NLG, the characteristics of UDGs have not been explored. Therefore, in this section, we consider three specific UDGs that have well-constrained baryonic mass profiles, namely, AGC 114905, 242019 and 2019533.  In Fig. \ref{AGC114905}, the first row shows the star and gas surface densities in these galaxies. Their observed rotation curves and the analytic curves obtained in NLG and MOND are shown in the second row. The total matter density and the effective dark matter density for the three galaxies are given in the third row. The best fit rotation curves in NLG are obtained by the Monte Carlo Markov Chain (MCMC) simulations and the corresponding likelihood corner plots are shown in the last row. We now discuss these galaxies in detail. 

\subsection{AGC 114905}
\begin{figure}
\centering

\includegraphics[width=4cm]{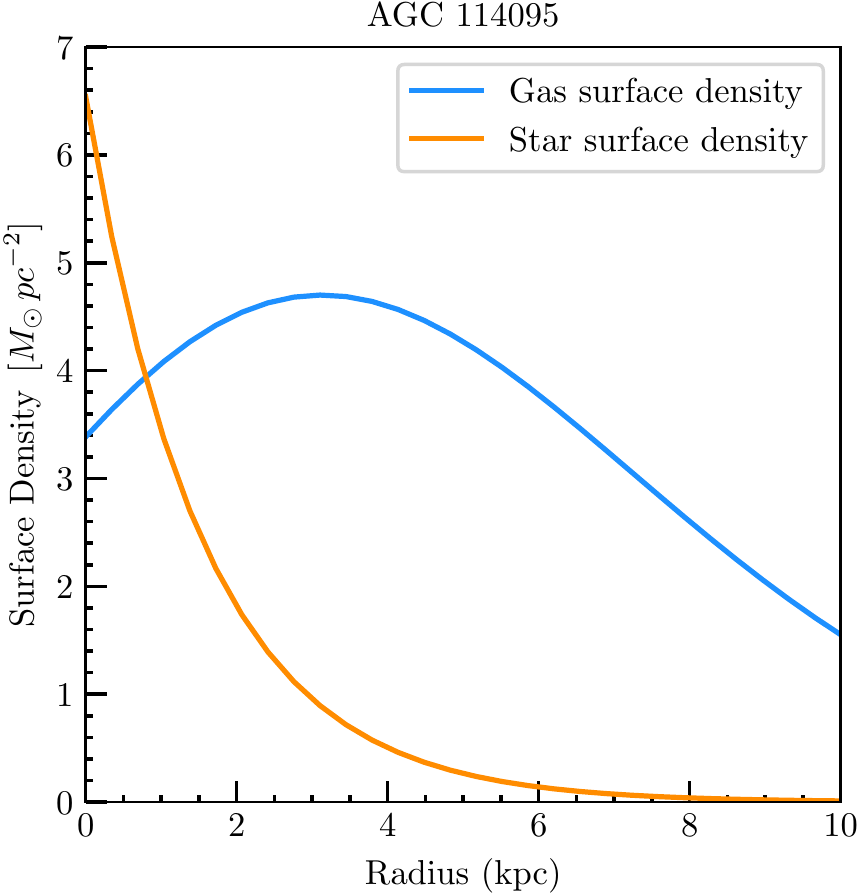}
\includegraphics[width=4.2cm]{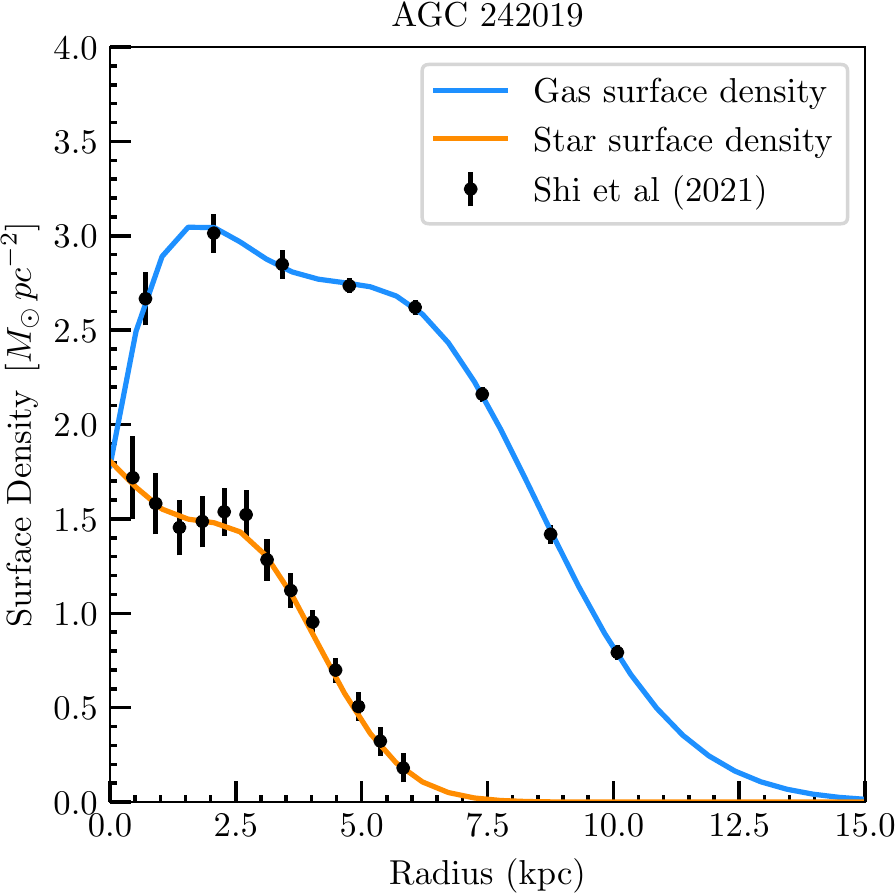}
\includegraphics[width=4cm]{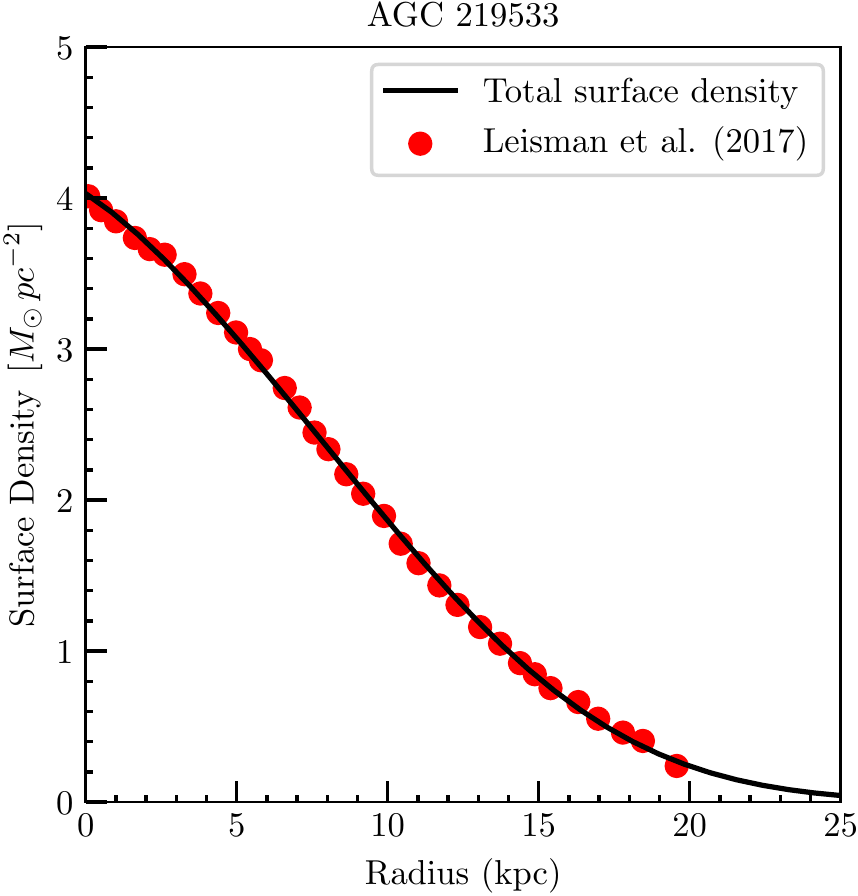}\vspace{0.3cm}
\includegraphics[width=4.0cm]{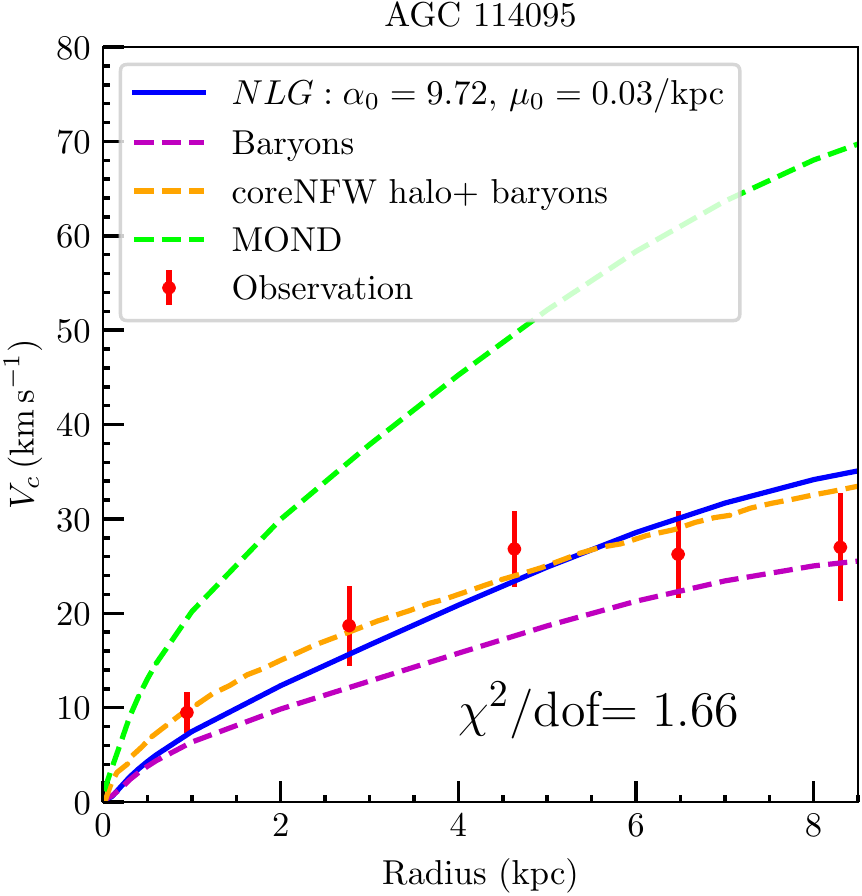}
\includegraphics[width=4.1cm]{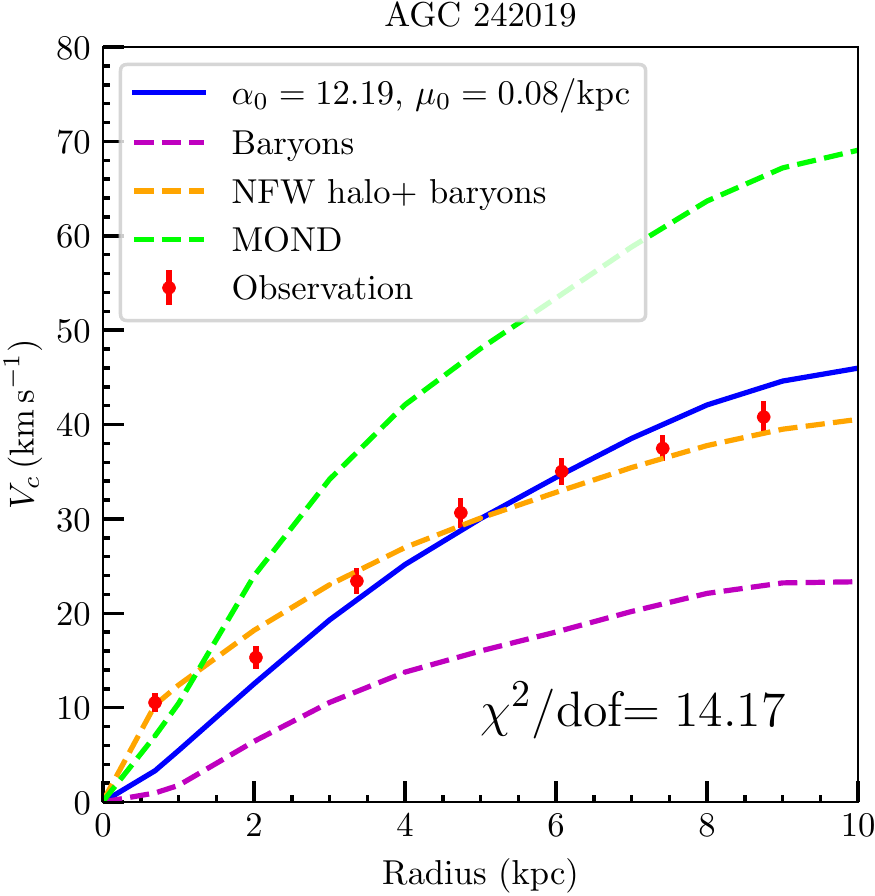}
\includegraphics[width=4.0cm]{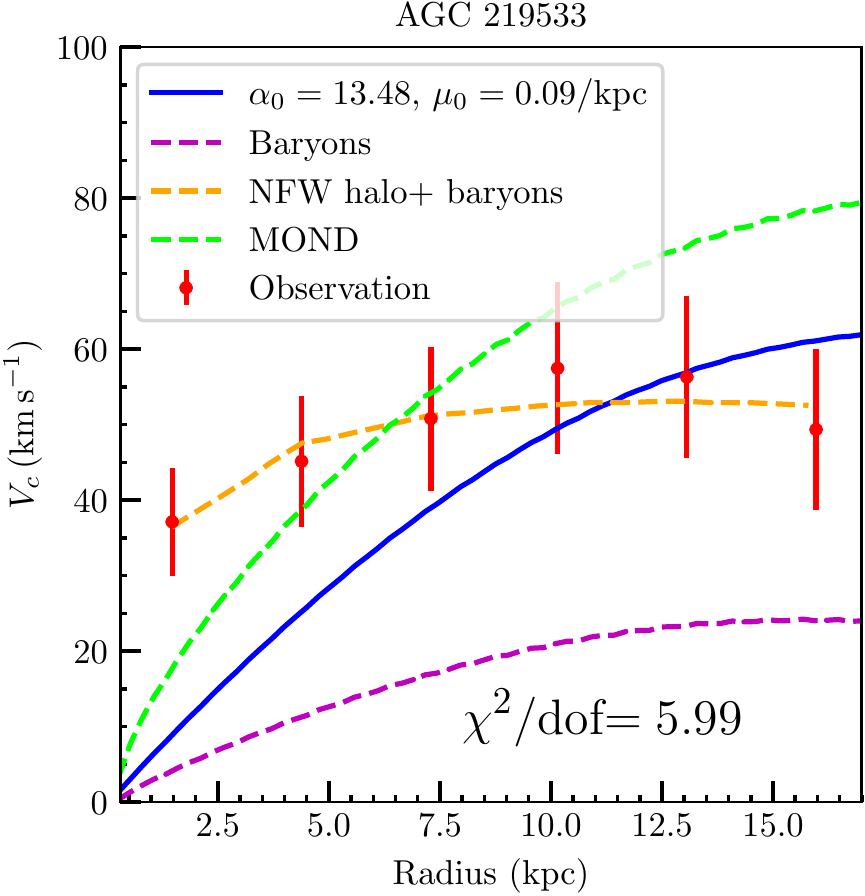}\vspace{0.3cm}
\includegraphics[width=4cm]{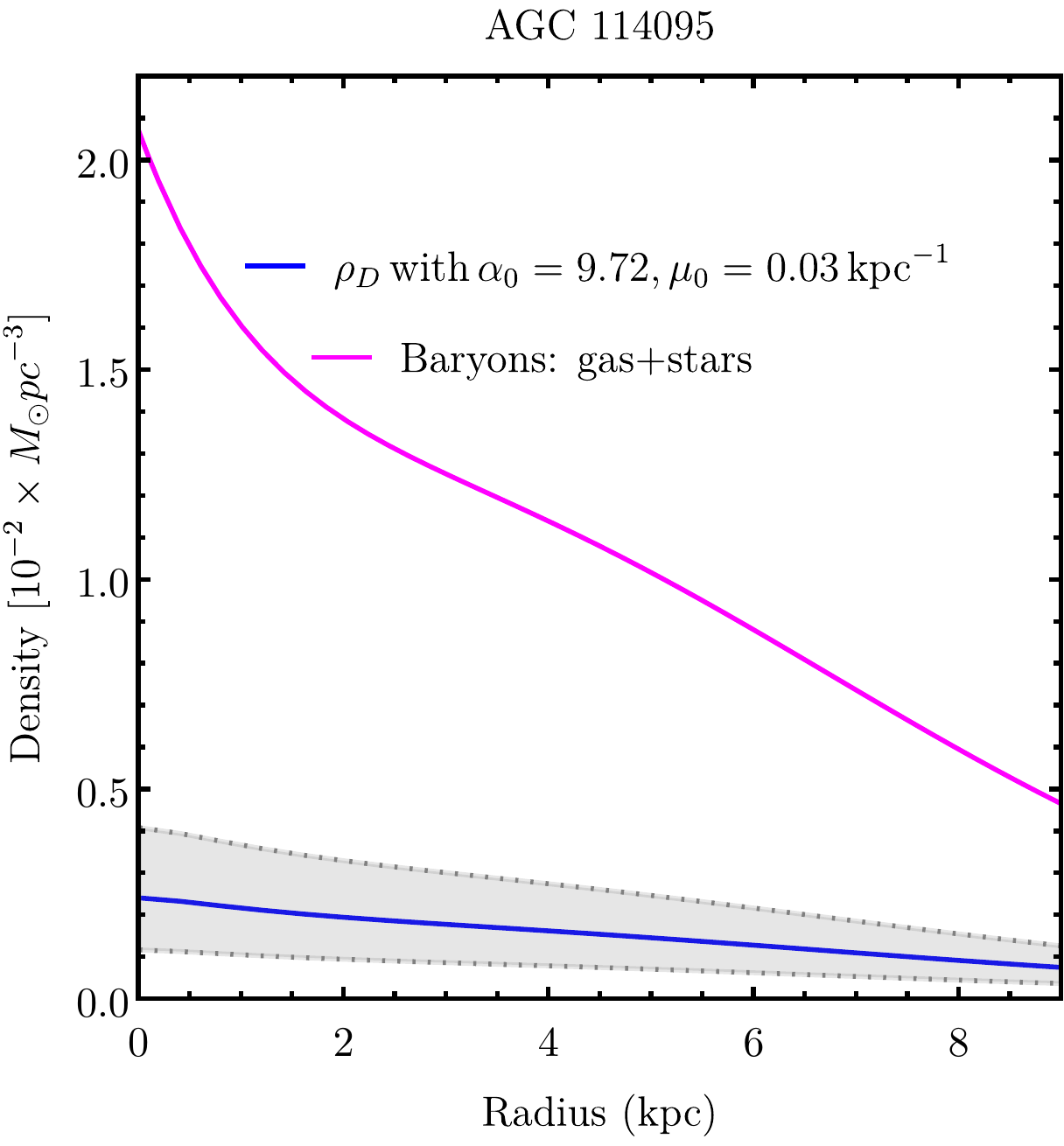}
\includegraphics[width=4cm]{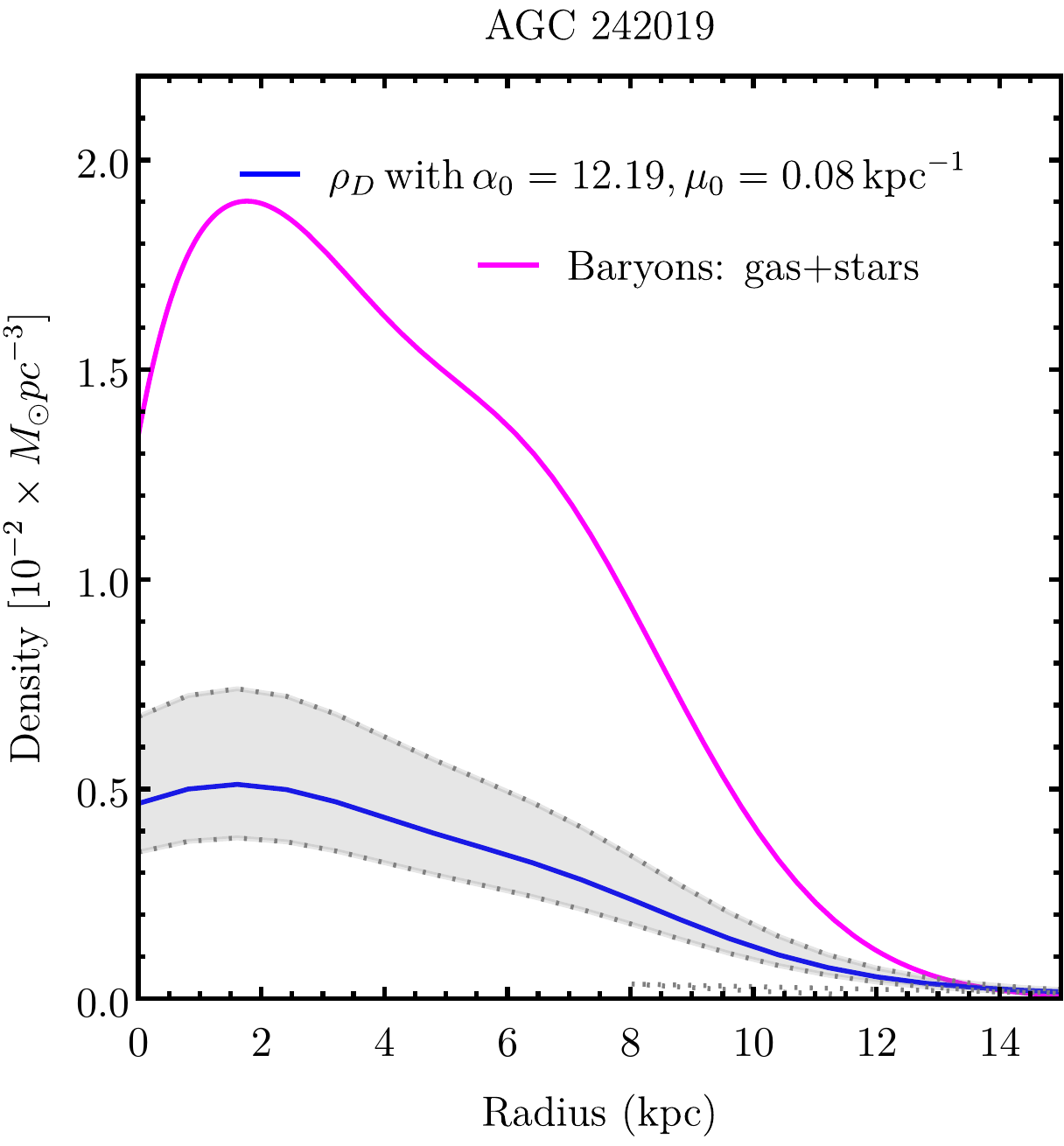}
\includegraphics[width=4cm]{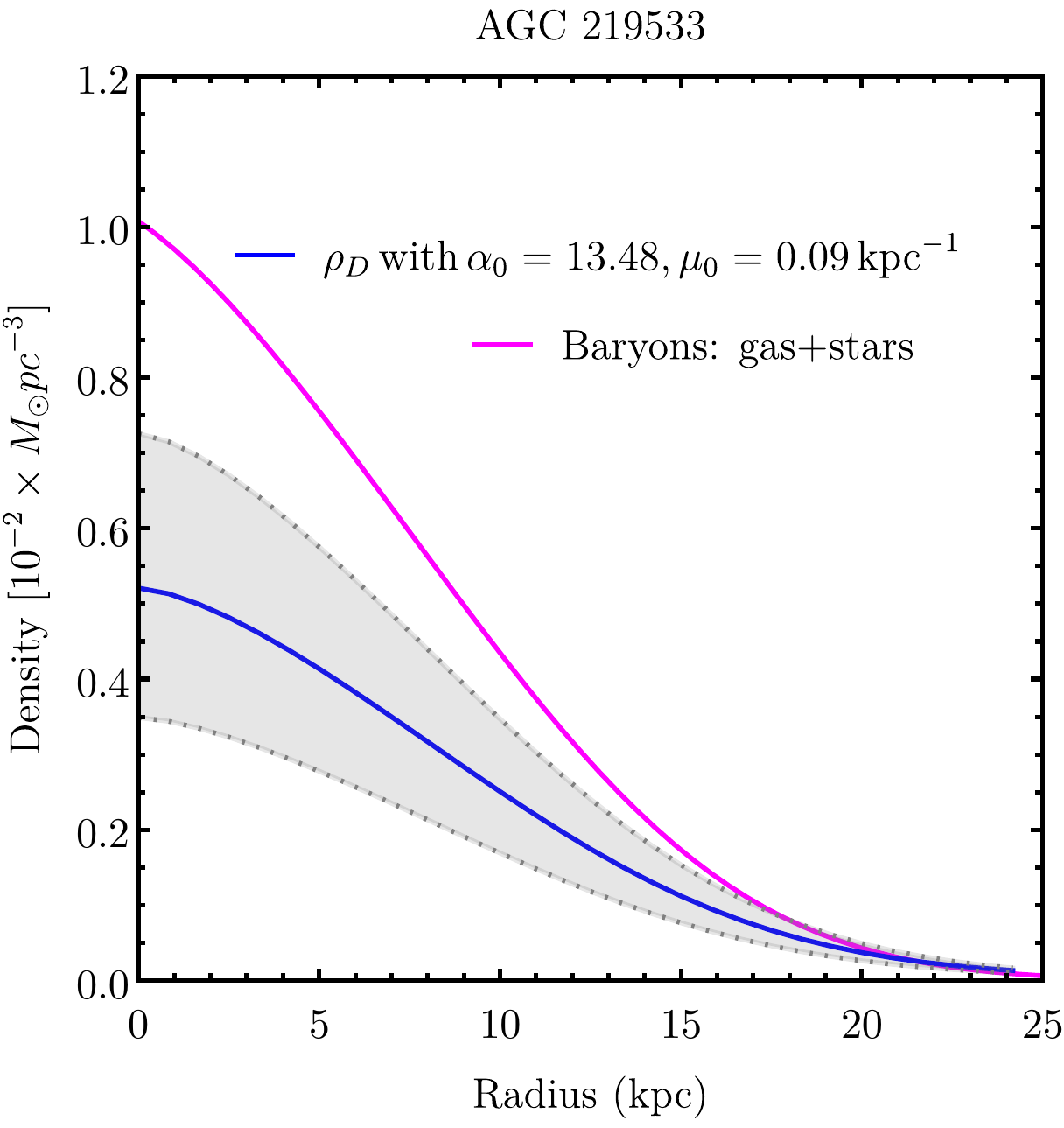}\vspace{0.3cm}
\includegraphics[width=4.2cm]{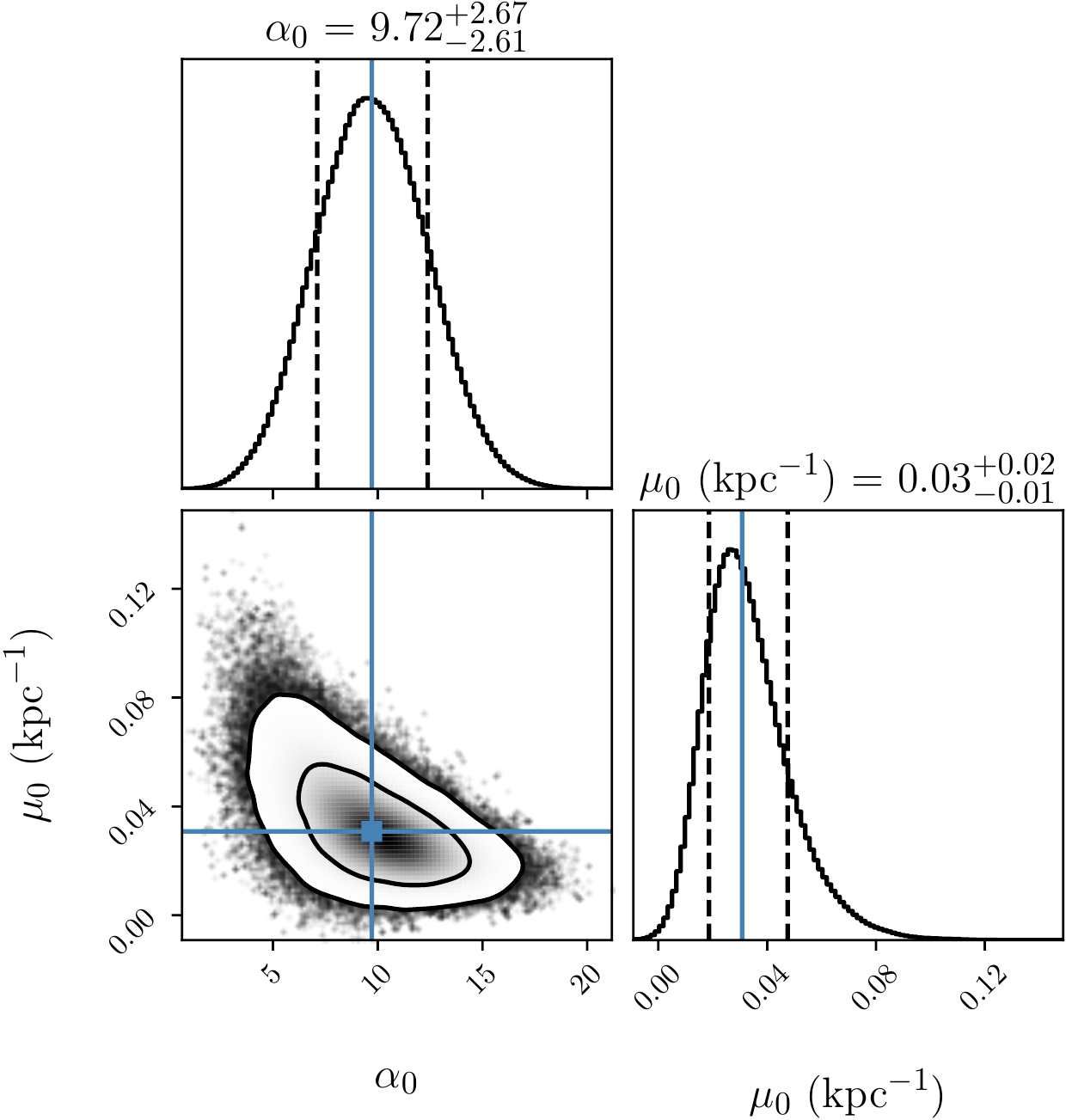}
\includegraphics[width=4.2cm]{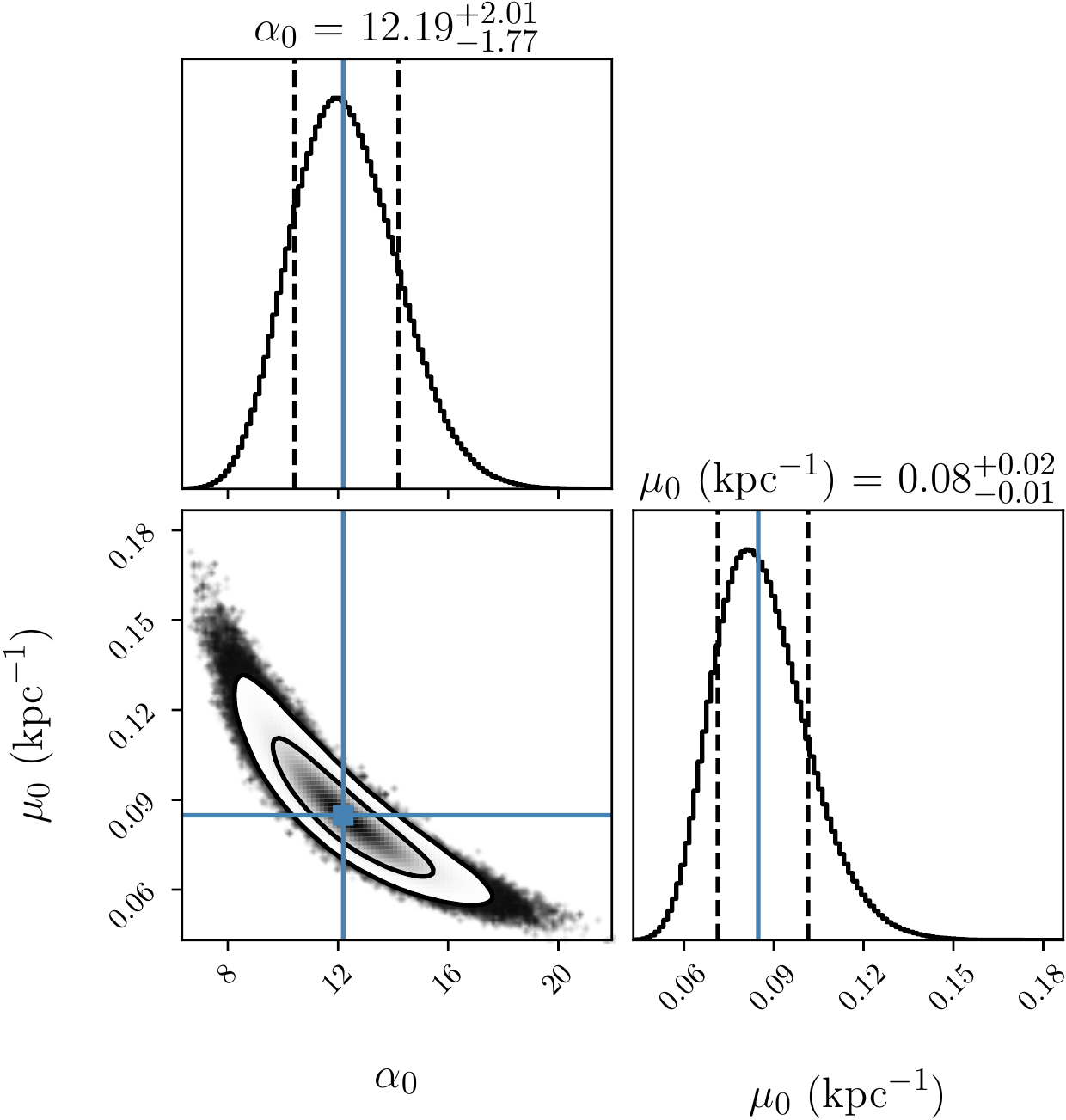}
\includegraphics[width=4.2cm]{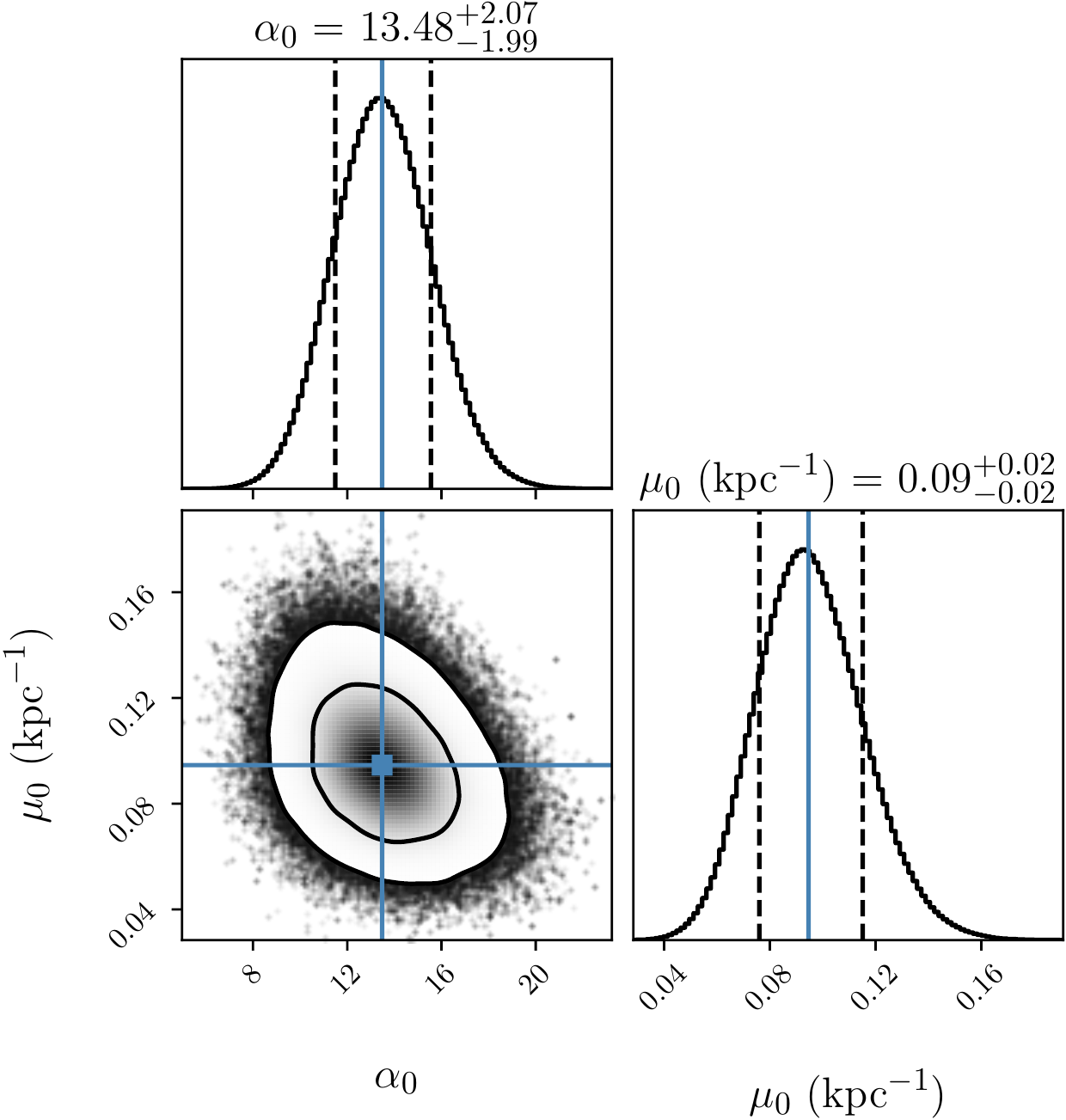}
\caption {The first row shows the baryonic surface density for each galaxy. The second row shows the rotation curves of the galaxies; the MOND rotation curve is calculated using the quasi-linear formulation known as QUMOND. The third row shows $\rho(R, z = 0)$, the total baryonic mass density, and $\rho_D(R, z = 0)$, the associated effective dark matter density according to NLG for the galactic disk ($z = 0$); for the latter curves, the gray regions enclosed between dotted curves in the panels reflect the $1\, \sigma$ bounds on best fit values of $\alpha_0$ and $\mu_0$ obtained by the {\tt emcee} code. Parameter corner plot for each galaxy is given in the last row; here, the posterior PDFs of the parameters are produced from MCMC simulations via  {\tt emcee} code.}\label{AGC114905} 
\end{figure}

As a recent example, let us first discuss AGC 114905, which is a gas-rich UDG with total baryonic mass $\mathcal{M}=(1.4\pm 0.2)\times 10^9\, M_{\odot}$. Its rotation curve seems to be flat and approaches $ v_c\approx 28\, \text{km}/\text{s}$ at radius $r\approx 8 \,$kpc. The recent high-resolution interferometric observations imply that this UDG does not follow the standard concentration-halo mass relation~\cite{PinaMancera:2021wpc}. More specifically, the dark matter content of the galaxy is much lower than expected.  Although it seems that there are some uncertainties in the determination of the inclination angle and the distance of this galaxy, it nevertheless poses a challenge for MOND~\cite{PinaMancera:2021wpc}. To be specific, one may assign a characteristic acceleration to this galaxy given by $v_c^2/r\approx 3.2\times 10^{-12}\, \text{m}/\text{s}^2$ that is much smaller than the MOND acceleration scale $\approx1.2\times 10^{-10}\, \text{m}/\text{s}^2$. In other words, this galaxy should lie within the deep MOND regime implying that the observed rotation curve should be much higher than the Newtonian rotation curve of the total baryonic mass. However, as mentioned, this is not the case and 
the observed rotation curve deviates strongly, at all radii, from the MONDian rotation curve.

To calculate within the Newtonian regime of NLG the rotation curve $V(R)$ for a spiral  galaxy, where $R$ is the distance from the center of the disk, we assume that the centripetal acceleration in circular motion within the disk is equal to gravitational attraction, namely,  $V^2/R = \mathcal{F}_{\rm NLG}$, where  $\mathcal{F}_{\rm NLG}$, the magnitude of the gravitational force per unit mass in the disk of the galaxy, is obtained via the integration of Eq.~\eqref{G2a} over the disk of the galaxy.  Let the origin of Cartesian coordinates $(x, y, z)$ be the center of the spiral galaxy. The disk is defined by the $z=0$ plane,  where we introduce polar coordinates $x = r \cos \phi$ and $y = r \sin \phi$. Here,  $\phi$ is the azimuthal coordinate $0 \to 2\pi$. With no loss in generality, we can assume in extending Eq.~\eqref{G2a} over the disk that $\mathbf{x}' = (R, 0, 0)$ and 
$\mathbf{x}_p = (r\cos \phi, r \sin \phi, 0)$ in our Cartesian coordinates and $U := |\mathbf{x}' - \mathbf{x}_p|$ is given by
\begin{equation}\label{Ua}
 U = (R^2 -2 Rr\cos \phi +r^2)^{1/2}\,.
\end{equation}
In Eq.~\eqref{G2a}, we replace $q$ by $q_0$ for the sake of simplicity  and note that
\begin{equation}\label{Ub}
 \int_0^U 4\,\pi\,s^2 q_0(s)\, ds = \alpha_0\,[1- (1+\tfrac{1}{2}\,\mu_0U)\,e^{-\mu_0U}]\,;
\end{equation}
moreover, replacing $m_p$ by $\Sigma(r)\, r \,dr d\phi$, where $\Sigma$ is the isotropic column density of baryonic matter in the disk, we find
\begin{equation}\label{Uc}
 \mathcal{F}_{\rm NLG} (R) = G\int _0^\infty \Sigma(r) r\, dr \int_0^{2\pi} \frac{R-r\cos\phi}{U^3}\left\{1 +\alpha_0\,\left[1- (1+\tfrac{1}{2}\,\mu_0U)\,e^{-\mu_0U}\right]\right\}\,d\phi\,.
\end{equation}

The main purpose of this subsection is to explore the rotation curve and the effective dark matter density of AGC 114905 in accordance with NLG. To do so, we need the baryonic mass distribution of the galaxy. We use the physical parameters reported in Figure 6 of Ref.~\cite{PinaMancera:2021wpc}. Let us briefly review the baryonic matter properties of AGC 114905. For the stellar disk, we use an exponential disk model with total stellar mass $M_{*}\approx 1.0\times 10^8\, M_{\odot}$ and radial scale length $R_d\approx 1.56\,$kpc. On the other hand, the vertical structure is governed by a $\text{sech}^2$ profile with scale height $z_d\approx 260\,$pc. Moreover, the surface density profile of the gas component is given by \cite{PinaMancera:2021wpc}
\begin{equation}  \label{U1}
\Sigma(R)=\Sigma_0\, e^{-\frac{R}{R_1}}\left(1+\frac{R}{R_2}\right)^{\beta}\,,
\end{equation}
where $R_1\approx 0.97\,$kpc, $R_2\approx 14.37\,$kpc, $\beta=18.04$ and $\Sigma_0\approx 3.39 \,M_{\odot}/\text{pc}^2$. The corresponding vertical distribution is given by a Gaussian profile with a constant vertical scale height $z_d \approx 250\,$pc. With these parameters, the total mass of the gas component including H \textsc{i} and Helium contributions is $M_{\text{gas}}\approx 1.345\times 10^9\, M_{\odot}$. The left panel in the first row of Fig. \ref{AGC114905} shows the gas and star surface densities of AGC 114905. 

The observational data regarding the rotation curve of AGC 114905 are shown in the left panel of the second row of Fig.~\ref{AGC114905}. The observed rotation curve and the standard dark matter halo model rotation curve have been recovered from Figure 6 of Ref.~\cite{PinaMancera:2021wpc}. The MONDian rotation curve is obtained using the QUMOND formulation \cite{PinaMancera:2021wpc}. The baryonic (without dark matter) rotation curve is derived using the baryonic matter surface density. The blue curve indicates the rotation curve in accordance with NLG. To obtain this curve, we use the public code {\tt emcee} \cite{Foreman-Mackey:2012any} to obtain the best fit values for $\alpha_0$ and $\mu_0$. To do so, the Gaussian distribution is assumed as prior constraint on both parameters, see Table \ref{tab1}. It should be noted that the main observational values of the parameters as well as the error bars have been used as the mean and standard deviation of the prior Gaussian distribution, respectively. The posterior probability density functions (PDFs) of the parameters are derived from MCMC simulations through {\tt emcee} code. In our analysis, the posterior probability of parameters $\boldsymbol{\theta}=\{\theta_k\}$ is defined by
   \begin{equation}
     p(\boldsymbol{\theta}) \propto \exp\left( -\frac{\chi^2}{2} \right) \prod_k {\rm Pr}(\theta_k),
     \label{eq:pdf}
   \end{equation}
where ${\rm Pr}(\theta_k)$ is the prior probability of parameter $\theta_k$ and $\chi^2$ and is given by
   \begin{equation}
     \chi^2 = \sum_{j=1}^{N} \left[ \frac{v_{\rm c}(R_j) - V(\boldsymbol{\theta};R_j)}{\sigma_{v_{\rm c}(R_j)}} \right]^2,
     \label{eq:chi2}
   \end{equation}
 where ${v_{\rm c}(R_j)}$ is the observed circular velocity,  and $\sigma_{v_{\rm c}(R_j)}$ is its observed error. On the other hand, $V$ is the circular velocity in NLG. The degree of freedom, i.e., dof, is defined as the number of the data points minus the number of free parameters. The corresponding parameter corner plot is shown in the left panel of the last row in Fig. \ref{AGC114905}. The best fit value of the parameters and the corresponding $1\, \sigma$ values are $\alpha_0=9.72^{+2.67}_{-2.61}$ and $\mu_0=0.03^{+0.02}_{-0.01}\, \text{kpc}^{-1}$. These values are consistent with the observational constraints~\eqref{I9}. In this case, $\chi^2/\text{dof}=1.66$.
\begin{table}
\caption{Prior constraints on the model parameters. $\bar{\mu}$ and $\bar{\sigma}$ are the corresponding mean and standard deviation of the prior distribution. }
\begin{center}
  \begin{tabular}{lcc} \hline
 Parameter  &\,\,\,\,\,\,\,  Distribution  &\,\,\,\,\,\,\, ($\bar{\mu}$, $\bar{\sigma}$) \\
 \hline
 $\alpha_0$     & Gaussian  & ($10.94$, $2.56$)     \\
 $\mu_0$      & Gaussian  & ($0.059$, $0.028$)  \\
 \hline
\end{tabular}
\end{center}
\label{tab1}
\end{table}

It is clear that there is considerable deviation between this specific observational result and the MOND scenario. On the other hand, the rotation curve in accordance with NLG shows much better compatibility with the data. As we already mentioned, NLG predicts less dark matter in dwarf galaxies. This is why the NLG rotation curve does not deviate strongly from the baryonic contribution in AGC 114905. This fact can be seen in the left panel in the third row of Fig.~\ref{AGC114905}, where the effective dark matter density according to NLG is shown at the plane of the galaxy, i.e., $z=0$. The gray region indicates the $1\,\sigma$ region. We see that the effective dark matter density is much smaller than the baryonic mass density. In contrast, in the MOND scenario, this galaxy lies within the deep MONDian regime and must contain a large amount of phantom dark matter.   

It proves useful to compute the ratio $M_D/M$ within the sphere of radius $8.3\,$kpc. The result is
\begin{equation}    \label{U2}
\frac{M_D}{M}\approx 1.15^{+1.25}_{-0.59}\,.
\end{equation}
The error bar corresponds to the $1\, \sigma$ bounds on $\alpha_0$ and $\mu_0$ obtained by the {\tt emcee} code. This result is consistent with Eq.~\eqref{G8}. Furthermore, it is more or less in agreement with the cored cold dark matter model (\textsc{core}NFW) employed in \cite{PinaMancera:2021wpc}. The \textsc{core}NFW halo profile is an extension of the NFW profile; in this connection,  see Eq. (5) of Ref.~\cite{PinaMancera:2021wpc}, where ``Case 2" has been introduced as a suitable halo model for describing the rotation curve of AGC 114905. Using the fitting parameters introduced in~\cite{PinaMancera:2021wpc}, it is straightforward to show that within the sphere of radius $8.3\,$kpc, we have
\begin{equation}    \label{U3}
\frac{M_D}{M}\approx 0.68^{+4.19}_{-0.45}\,.
\end{equation}
The upper and lower limits correspond to the allowed range for the concentration parameter of the halo, i.e. $c_{200}=0.3^{+0.3}_{-0.2}$. It is necessary to mention that the concentration parameter is extremely low for this halo model and violates the so-called concentration-halo mass relation~\cite{Dutton:2014xda}. Therefore, although the formation of such a halo is beyond $\Lambda$CDM expectations, this type of galaxy can naturally exist in NLG. 

Finally, we emphasize that there are some uncertainties in the inclination angle and the distance of AGC 114905;  in  this connection, see~\cite{SS} for a recent study. For these parameters, we have used  the values given by the best fit model developed in~\cite{PinaMancera:2021wpc} via a Markov chain Monte Carlo routine.


\subsection{AGC 242019}
AGC 242019 is a UDG that is believed to host a cuspy dark matter halo \cite{Shi:2021tyg}. As we already discussed, NLG does not lead to cuspy effective dark matter halos. Therefore, it is natural to expect  that there would be conflict between NLG and  observational data for AGC 242019. The observed gas and star surface densities of AGC 242019 are shown in the middle panel of the first row in Fig. \ref{AGC114905}. To calculate the effective dark matter density, we need analytic expressions for baryonic matter densities. Therefore we have fitted the following functions to the surface density data
\begin{equation}
\Sigma_{\text{star}}(R)= \Big(1-\frac{R}{R_1}+\frac{R^2}{R_2^2}+\frac{R^3}{R_3^3}\Big)\Sigma_0^*\, e^{-\frac{R^2}{R_4^2}}\,
\end{equation}
and
\begin{equation}
\Sigma_{\text{gas}}(R)= \Big(1+\frac{R}{\bar{R}_1}-\frac{R^2}{\bar{R}_2^2}+\frac{R^3}{\bar{R}_3^3}\Big)\Sigma_0\, e^{-\frac{R^2}{\bar{R}_4^2}}\,.
\end{equation}
We use the {\tt emcee} code to obtain the best fit value of the parameters. By assuming a uniform prior distribution for all the coefficients $R_i$, $\bar{R}_i$, $\Sigma_0$ and $\Sigma_0^*$, we find $R_1=7.66\,$kpc, $R_2=3.70\,$kpc, $R_3=2.63\,$kpc, $R_4=2.63\,$kpc, $\Sigma_0^*=1.81\,M_{\odot}/\text{pc}^2$, $\bar{R}_1=1.04\,$kpc, $\bar{R}_2=1.72\,$kpc, $\bar{R}_3=2.59\,$kpc, $\bar{R}_4=4.82\,$kpc and $\Sigma_0=1.78\, M_{\odot}/\text{pc}^2$. The resulting curves are shown with orange and blue solid curves in the middle panel of the top row in Fig. \ref{AGC114905}. For the vertical structure of the disks, we take the sech$^2$ profile with scale height $0.2\,$kpc for the stellar disk and $0.1\,$kpc for the gas disk \cite{Shi:2021tyg}. Eventually, the NLG rotation curve is found using the same procedure described for the AGC 114905 galaxy. The result is shown in the middle panel of the second row in Fig. \ref{AGC114905}. On the other hand, the parameter corner plot is shown in the middle panel of the last row indicating that $\alpha_0=12.19^{+2.01}_{-1.77}$ and $\mu_0=0.08^{+0.02}_{-0.01}\, \text{kpc}^{-1}$. These values are consistent with the observational constraints \eqref{I9}. For NLG,  the large $\chi^2/\text{dof}=14.17$ is due to the innermost data point close to the center; indeed, ignoring this data point and repeating the MCMC simulations, we find $\chi^2/\text{dof}= 3.78$.  The effective dark matter distribution is shown in the middle panel of the third row. Finally, let us mention that the MONDian rotation curve is shown in the middle panel of the second row. Clearly, MOND's rotation curve disagrees with the observational data of AGC 242019 \cite{Shi:2021tyg}.

\subsection{AGC 219533}

Among UDGs studied in \cite{LHJ}, AGC 219533 has sufficient observational data to be explored here. More specifically, its observed rotation curve and the total surface density are known \cite{LHJ}. The right panel of the first row of Fig. \ref{AGC114905} shows the total surface density of the galaxy. The black curve is obtained using {\tt emcee} code by finding the best fit values of $R_1$, $R_2$, $R_3$ and $\Sigma_0$ in the following function
\begin{equation}
\Sigma(R)= \Big(1-\frac{R}{R_1}+\frac{R^2}{R_2^2}\Big)\Sigma_0 \,e^{-\frac{R^2}{R_3^2}}\,;
\end{equation}
indeed, the best fit values are $R_1=31\,$kpc, $R_2=14.20\,$kpc, $R_3=10\,$kpc, $\Sigma_0= 4.03\,M_{\odot}/\text{pc}^2$. For the vertical distribution we take the sech$^2$ profile with scale height $0.2\,$kpc. Having specified the baryonic matter distribution, we find the best fit rotation curve using the same procedure outlined for AGC 114905. The result alongside the MONDian rotation curve is shown in the right panel of the second row. The baryonic rotation curve is shown with the purple dashed curve.
In this case, the best fit  parameter values are $\alpha_0=13.48^{+2.07}_{-1.99}$ and $\mu_0=0.09^{+0.02}_{-0.02}\, \text{kpc}^{-1}$ and  the distribution of the effective dark matter is shown in the right panel of the third row. For AGC 219533, we find $\chi^2/\text{dof}=5.99$, which confirms that NLG does not give a suitable fit. 
However, it is worth noting that there is a serious limitation to the validity of the observational data of this galaxy. The analysis done in \citep{LHJ} suffers from a lack of careful kinematic modeling for the rotation curve and the surface densities. It is shown in \citep{Pina2020} that the rotation curve and also the gas surface density profile can have at best two resolution elements per galaxy side. In other words, the observed beam is too wide to fit more points. Therefore, any data set including more than two points are completely correlated, and cannot be fully trusted. Naturally, the two data points reported in \cite{Pina2020} are not enough to make an accurate MCMC fitting analysis; hence, we decided to use \citep{LHJ} data and at the same time stress its limitations.

It would be interesting to find the mean values of the parameters $\alpha_0$ and $\mu_0$ from these three UDGs. The result is
\begin{equation}
\alpha_0=11.80^{+2.25}_{-2.12}\,, \qquad  \mu_0=0.07^{+0.02}_{-0.01}\, \text{kpc}^{-1}\,,
\end{equation}
although these values are consistent with those obtained from normal spiral galaxies, it is clear that, based on these three UDGs, somewhat larger values for the free parameters $(\alpha_0,\mu_0)$ are required in order to explain the properties of UDGs.

\section{Discussion}

In the Newtonian regime of NLG, the nonlocal aspect of the universal gravitational interaction appears as an extra density of matter $\rho_D(\mathbf{x})$ in the Poisson equation for the Newtonian gravitational potential. Moreover, $\rho_D$ is the convolution of matter density with an empirically determined kernel, which is assumed to be a smooth positive spherically symmetric function that is independent of the density of matter.  It is natural to interpret $\rho_D$ as the density of effective dark matter. Compared to the standard particle dark matter model, two features of this modified gravity model stand out: $\rho_D(\mathbf{x})$ is in general a smooth cusp-free function and, furthermore, the net amount of effective dark matter in a dwarf galaxy is less than what is expected in accordance with the dark matter paradigm. We compare these implications of NLG with observation. In connection with the former feature, we note that  it is not possible to deduce $\rho_D(0)$ from the observed rotation curves; hence, this feature is not in conflict with observation. Regarding the latter feature, we have studied the rotation curves of three UDGs, namely,  AGC 114905, 242019 and 219533. It has tuned out that NLG naturally predicts rotation results close to the data of the three dwarf galaxies. This tentative conclusion is based on fitting the  observational data reasonably well for AGC 114905, but not as well for the other two galaxies. This situation with NLG is in sharp contrast with the standard cold dark matter paradigm, where the rotation curves can be fit very well; however, the resulting predicted total amount of dark matter for each UDG is too small by  about an order of magnitude. This shortcoming, on the other hand, could be the result of rare collision events recently proposed by van Dokkum \textit{et al.}~\cite{vanDokkum:2022zdd}.

\section*{Acknowledgments}
 
The work of M. R. has been supported by the Ferdowsi University of Mashhad. Moreover, M. R. is grateful to Neda Ghafourian and Tahere Kashfi for useful discussions and suggestions in connection with  Figure~\ref{AGC114905}. The authors wish to thank Pavel E. Mancera~Pi\~na for helpful discussions and for kindly sharing the observational data regarding the baryonic content of AGC 114905. The authors are also grateful to the referee for comments and suggestions that helped to improve this paper.

\appendix


\begin{thebibliography}{99}

\bibitem{FZ}
F. Zwicky,
``Die Rotverschiebung von extragalaktischen Nebeln",
Helv. Phys. Acta {\bf 6}, 110-127 (1933). 
[English translation: F. Zwicky,
 ``The redshift of extragalactic nebulae", 
 Gen. Relativ. Gravit. {\bf 41}, 207-224 (2009)].
 
\bibitem{Jackson}
J. D. Jackson, \emph{Classical Electrodynamics}, 3rd edn (Wiley, Hoboken, NJ, 1999). 

\bibitem{L+L}
L. D. Landau and E. M. Lifshitz,
\emph{Electrodynamics of Continuous Media} (Pergamon, Oxford, UK, 1960).


\bibitem{HeOb}  
F. W. Hehl, and Y. N. Obukhov, 
\emph{Foundations of Classical Electrodynamics: Charge, Flux, and Metric} 
(Birkh\"auser, Boston, MA, USA, 2003). 

\bibitem{Einstein}
A. Einstein, 
\emph{The Meaning of Relativity} 
(Princeton University Press, Princeton, NJ, USA, 1955).


\bibitem{BMB}
B. Mashhoon,
\emph{Nonlocal Gravity}
(Oxford University Press, Oxford, UK, 2017).



\bibitem{Cho}
Y. M. Cho, 
``Einstein Lagrangian as the translational Yang-Mills Lagrangian", 
Phys. Rev. D {\bf 14}, 2521-2525 (1976). 



\bibitem{Hehl:2008eu}
F.~W.~Hehl and B.~Mashhoon,
``Nonlocal Gravity Simulates Dark Matter",
Phys. Lett. B \textbf{673}, 279-282 (2009).
[arXiv:0812.1059 [gr-qc]]

              

\bibitem{Hehl:2009es}
F.~W.~Hehl and B.~Mashhoon,
``Formal framework for a nonlocal generalization of Einstein's theory of gravitation",
Phys. Rev. D \textbf{79}, 064028 (2009).
[arXiv:0902.0560 [gr-qc]]
 
\bibitem{Bini:2016phe}
D.~Bini and B.~Mashhoon,
``Nonlocal gravity: Conformally flat spacetimes",
Int. J. Geom. Meth. Mod. Phys. \textbf{13}, no.06, 1650081 (2016).
[arXiv:1603.09477 [gr-qc]]

\bibitem{Chicone:2011me}
C.~Chicone and B.~Mashhoon,
``Nonlocal Gravity: Modified Poisson's Equation",
J. Math. Phys. \textbf{53}, 042501 (2012).
[arXiv:1111.4702 [gr-qc]]

\bibitem{Toh}
J. E. Tohline,
``Does gravity exhibit a $1/r$ force on the scale of galaxies?", 
Ann. N.Y. Acad. Sci. {\bf 422}, 390 (1984).

\bibitem{Kuhn}
J. R. Kuhn and L. Kruglyak, 
``Non-Newtonian forces and the invisible mass problem", 
Astrophys. J. {\bf 313}, 1-12 (1987). 

\bibitem{Bek}
J. D. Bekenstein, 
``The missing light puzzle: a hint about gravity?",  in \emph{Second Canadian Conference on General Relativity and Relativistic Astrophysics}, A. Coley, C. Dyer, and T. Tupper, eds. (World Scientific, Singapore, 1988), pp. 68-104.

\bibitem{Chicone:2015coa}
C.~Chicone and B.~Mashhoon,
``Nonlocal Gravity in the Solar System",
Classical Quantum Gravity \textbf{33}, no.7, 075005 (2016).
[arXiv:1508.01508 [gr-qc]]
 

\bibitem{Rahvar:2014yta}
S.~Rahvar and B.~Mashhoon,
``Observational Tests of Nonlocal Gravity: Galaxy Rotation Curves and Clusters of Galaxies",
Phys. Rev. D \textbf{89}, 104011 (2014).
[arXiv:1401.4819 [astro-ph.GA]]

\bibitem{A+S}
M. Abramowitz and I. A. Stegun, 
\emph{Handbook of Mathematical Functions} (National Bureau of Standards, Washington, D.C., 1964). 



\bibitem{Plummer}
H. C. Plummer,
``On the Problem of Distribution in Globular Star Clusters", 
Mon. Not. Roy. Astron. Soc. \textbf{71}, 460-470 (1911).

\bibitem{Bog}
V. I. Bogachev, \emph{Measure Theory}, vol. I (Springer-Verlag, Belin, 2007). 


\bibitem{Binney}
J.~Binney and S.~Tremaine, \textit{Galactic Dynamics}, 2nd edn (Princeton University Press, Princeton, NJ, USA, 2008).
  


 \bibitem{Boldrini:2019yvk}
P.~Boldrini, R.~Mohayaee and J.~Silk,
``Fornax globular cluster distributions: implications for the cusp-core problem",
Mon. Not. Roy. Astron. Soc. \textbf{485}, no.2, 2546-2557 (2019).
[arXiv:1903.00354 [astro-ph.GA]]



\bibitem{NFW}
J. F. Navarro, C. S. Frenk and S. D. M. White, 
``The Structure of Cold Dark Matter Halos", 
Astrophys. J. \textbf{462}, 563-575 (1996).
[arXiv: astro-ph/9508025]



\bibitem{universal}
M.~G.~Walker, M.~Mateo, E.~W.~Olszewski, J.~Penarrubia, N.~W.~Evans and G.~Gilmore,
``A Universal Mass Profile for Dwarf Spheroidal Galaxies",
Astrophys. J. \textbf{704}, 1274-1287 (2009)
[erratum: Astrophys. J. \textbf{710}, 886-890 (2010)].
[arXiv:0906.0341 [astro-ph.CO]]


\bibitem{McGaugh:2021tyj}
S.~S.~McGaugh, F.~Lelli, J.~M.~Schombert, P.~Li, T.~Visgaitis, K.~S.~Parker and M.~S.~Pawlowski,
``The Baryonic Tully\textendash{}Fisher Relation in the Local Group and the Equivalent Circular Velocity of Pressure-supported Dwarfs",
Astron. J. \textbf{162}, no.5, 202 (2021).
[arXiv:2109.03251 [astro-ph.GA]]









\bibitem{vanDokkum:2018vup}
P.~van Dokkum, S.~Danieli, Y.~Cohen, A.~Merritt, A.~J.~Romanowsky, R.~Abraham, J.~Brodie, C.~Conroy, D.~Lokhorst and L.~Mowla, \textit{et al.}
``A galaxy lacking dark matter",
Nature \textbf{555}, no.7698, 629-632 (2018).
[arXiv:1803.10237 [astro-ph.GA]]


\bibitem{Guo:2019wgb}
Q.~Guo, H.~Hu, Z.~Zheng, S.~Liao, W.~Du, S.~Mao, L.~Jiang, J.~Wang, Y.~Peng and L.~Gao, \textit{et al.}
``Further evidence for a population of dark-matter-deficient dwarf galaxies",
Nature Astron. \textbf{4}, no.3, 246-251 (2019).
[arXiv:1908.00046 [astro-ph.GA]]

\bibitem{Pina:2019rer}
P.~E.~Mancera Pi\~na, F.~Fraternali, E.~A.~K.~Adams, A.~Marasco, T.~Oosterloo, K.~A.~Oman, L.~Leisman, E.~M.~di Teodoro, L.~Posti and M.~Battipaglia, \textit{et al.}
``Off the Baryonic Tully\textendash{}Fisher Relation: A Population of Baryon-dominated Ultra-diffuse Galaxies",
Astrophys. J. Lett. \textbf{883}, no.2, L33 (2019).
[arXiv:1909.01363 [astro-ph.GA]]

\bibitem{Hammer:2020qcd}
F.~Hammer, Y.~Yang, F.~Arenou, J.~Wang, H.~Li, P.~Bonifacio and C.~Babusiaux,
``Orbital evidences for dark-matter-free Milky Way dwarf spheroidal galaxies",
Astrophys. J. \textbf{892}, no.1, 3 (2020).
[arXiv:2002.09493 [astro-ph.GA]]



\bibitem{Shen:2021zka}
Z.~Shen, S.~Danieli, P.~van Dokkum, R.~Abraham, J.~P.~Brodie, C.~Conroy, A.~E.~Dolphin, A.~J.~Romanowsky, J.~M.~Diederik Kruijssen and D.~Dutta Chowdhury,
``A Tip of the Red Giant Branch Distance of 22.1 \ensuremath{\pm} 1.2 Mpc to the Dark Matter Deficient Galaxy NGC 1052\textendash{}DF2 from 40 Orbits of Hubble Space Telescope Imaging",
Astrophys. J. Lett. \textbf{914}, no.1, L12 (2021).
[arXiv:2104.03319 [astro-ph.GA]]


\bibitem{Roshan:2021mfc}
M.~Roshan, I.~Banik, N.~Ghafourian, I.~Thies, B.~Famaey, E.~Asencio and P.~Kroupa,
``Barred spiral galaxies in modified gravity theories",
Mon. Not. Roy. Astron. Soc. \textbf{503}, no.2, 2833-2860 (2021).
[arXiv:2103.01794 [astro-ph.GA]]

\bibitem{Roshan:2021ljs}
M.~Roshan and B.~Mashhoon,
``Dynamical Friction in Nonlocal Gravity",
Astrophys. J. \textbf{922}, no.1, 9 (2021).
[arXiv:2107.05841 [gr-qc]]

\bibitem{McGaugh:2000sr}
S.~S.~McGaugh, J.~M.~Schombert, G.~D.~Bothun and W.~J.~G.~de Blok,
``The Baryonic Tully\textendash{}Fisher Relation",
Astrophys. J. Lett. \textbf{533}, L99-L102 (2000).
[arXiv:astro-ph/0003001 [astro-ph]]

\bibitem{vanDokkum:2022zdd}
P.~van Dokkum, Z.~Shen, M.~A.~Keim, S.~Trujillo-Gomez, S.~Danieli, D.~D.~Chowdhury, R.~Abraham, C.~Conroy, J.~M.~D.~Kruijssen and D.~Nagai, \textit{et al.}
``A trail of dark-matter-free galaxies from a bullet-dwarf collision",
Nature \textbf{605}, no.7910, 435-439 (2022).
[arXiv:2205.08552 [astro-ph.GA]]



\bibitem{PinaMancera:2021wpc}
P.~E.~Mancera~Pi\~na, F.~Fraternali, T.~Oosterloo, E.~A.~K.~Adams, K.~A.~Oman and L.~Leisman,
``No need for dark matter: resolved kinematics of the ultra-diffuse galaxy AGC 114905",
Mon. Not. Roy. Astron. Soc. \textbf{512}, no.3, 3230-3242 (2022).
[arXiv:2112.00017 [astro-ph.GA]]

\bibitem{Foreman-Mackey:2012any}
D.~Foreman-Mackey, D.~W.~Hogg, D.~Lang and J.~Goodman,
``emcee: The MCMC Hammer",
Publ. Astron. Soc. Pac. \textbf{125}, 306-312 (2013).
[arXiv:1202.3665 [astro-ph.IM]]

\bibitem{Dutton:2014xda}
A.~A.~Dutton and A.~V.~Macci\`o,
``Cold dark matter haloes in the Planck era: evolution of structural parameters for Einasto and NFW profiles",
Mon. Not. Roy. Astron. Soc. \textbf{441}, no.4, 3359-3374 (2014).
[arXiv:1402.7073 [astro-ph.CO]]

\bibitem{SS} 
J.~A.~Sellwood and R.~H.~Sanders,
``The ultra-diffuse galaxy AGC 114905 needs dark matter",
[arXiv:2202.08678 [astro-ph.GA]].


\bibitem{Shi:2021tyg}
Y.~Shi, Z.~Y.~Zhang, J.~Wang, J.~Chen, Q.~Gu, X.~Yu and S.~Li,
``A cuspy dark matter halo",
Astrophys. J. \textbf{909}, no.1, 20 (2021).
[arXiv:2101.01282 [astro-ph.GA]]




\bibitem{LHJ}
L.~Leisman, M.~P.~Haynes, S.~Janowiecki, et al.,
``(Almost) Dark Galaxies in the ALFALFA Survey: Isolated H \textsc{i}-bearing Ultra-diffuse Galaxies",
Astrophys. J. \textbf{842}, 133 (2017).



\bibitem{Pina2020}
P.~E.~Mancera Pi\~na, F.~Fraternali, K.~A.~Oman, E.~A.~K.~Adams, C.~Bacchini, A.~Marasco, T.~Oosterloo, G.~Pezzulli, L.~Posti and L.~Leisman, \textit{et al.}
``Robust H \textsc{i} kinematics of gas-rich ultra-diffuse galaxies: hints of a weak-feedback formation scenario",
Mon. Not. Roy. Astron. Soc. \textbf{495}, no.4, 3636-3655 (2020).
[arXiv:2004.14392 [astro-ph.GA]]

\end{thebibliography}
\end{document}